\documentclass[10pt,letterpaper]{article}
\usepackage[top=0.85in,left=2.75in,footskip=0.75in]{geometry}

\usepackage{amsmath,amssymb}

\usepackage{changepage}

\usepackage{textcomp,marvosym}

\usepackage{cite}

\usepackage{nameref,hyperref}

\usepackage[right]{lineno}

\usepackage[nopatch=eqnum]{microtype}
\DisableLigatures[f]{encoding = *, family = * }

\usepackage[table]{xcolor}

\usepackage{array}

\usepackage{xspace}

\newcolumntype{+}{!{\vrule width 2pt}}

\newlength\savedwidth

\raggedright
\newlength{\maxwidth}
\newlength{\maxheight}
\usepackage[font=footnotesize,hypcap=true,aboveskip=8pt,labelfont=bf,labelsep=period,justification=raggedright,singlelinecheck=off]{caption}

\makeatletter
\renewcommand{\@biblabel}[1]{\quad#1.}
\makeatother

\usepackage{lastpage,fancyhdr,graphicx}
\usepackage{epstopdf}
\fancyheadoffset[L]{2.25in}
\fancyfootoffset[L]{2.25in}
\usepackage{todonotes}
\reversemarginpar
\usepackage{csquotes}
\usepackage{float}
\usepackage{nth}

\usepackage{dejavu}
\usepackage{tgpagella}
\usepackage{mathpazo}
\usepackage[artemisia]{textgreek}

\usepackage{letltxmacro}
\LetLtxMacro\oldttfamily\ttfamily
\DeclareRobustCommand{\ttfamily}{\oldttfamily\csname ttsize\endcsname}
\newcommand{\setttsize}[1]{\def\ttsize{#1}}%
\setttsize{\footnotesize}

\usepackage{cleveref}
\crefname{table}{Table}{Tables}
\crefname{figure}{Fig.}{Figs.}
\crefname{section}{\S\hspace{-0.2ex}}{\S\S\hspace{-0.2ex}}
\crefname{equation}{}{}

\DeclareFontFamily{OT1}{pzc}{}
\DeclareFontShape{OT1}{pzc}{m}{it}{<-> s * [1.10] pzcmi7t}{}
\DeclareMathAlphabet{\mathpzc}{OT1}{pzc}{m}{it}

\usepackage[inline]{enumitem}
\setlist[enumerate]{label=(\arabic*), itemjoin={;\ }, itemjoin*={;\ and\ }}

\usepackage{url}

\hypersetup{pdfborder = 0 0 0}

\newlist{panels}{enumerate*}{3}
\setlist[panels]{label=\textbf{(\Alph*)}}

\newcommand*{\fig}{Fig~}

\newcommand*{\eqs}{Eqs.~}

\newcommand*{\ie}{\emph{i.e.}\xspace}
\newcommand*{\eg}{\emph{e.g.}\xspace}

\renewcommand{\phi}{\varphi}
\newcommand*{\ubar}{\overline{u}}
\renewcommand{\Pr}{\mathrm{Pr}}
\newcommand{\card}[1]{\left\vert{#1}\right\vert}

\newcommand*{\expectation}{\mathbb{E}}
\newcommand*{\given}{\mid}
\newcommand{\tpm}[1]{\mathcal{T}_{#1}}

\renewcommand{\Phi}{\mathrm{\emph{\textPhi}}}

\newcommand*{\phistructure}{\emph{\textPhi}-structure}
\newcommand*{\phistructures}{\emph{\textPhi}-structures}
\newcommand*{\phifold}{\emph{\textPhi}-fold}
\newcommand*{\phifolds}{\emph{\textPhi}-folds}

\newcommand*{\ces}{C}

\newcommand*{\environment}{E}
\newcommand*{\sensoryinterface}{\partial\system}
\newcommand*{\system}{S}
\newcommand*{\noiseenvironment}{N}

\newcommand*{\stimulus}{x}
\newcommand*{\responsevar}{Y}
\newcommand*{\responsestate}{y}
\newcommand*{\stimulusset}{\stimulus(1,k)}
\newcommand*{\responsestateset}{\responsestate(1,k)}
\newcommand*{\responsevarset}{\responsevar(1,k)}
\newcommand*{\noisestimulus}{\noiseenvironment}
\newcommand*{\noisestimulusset}{\noisestimulus(1,k)}
\newcommand*{\noiseresponsestate}{Y'}
\newcommand*{\noiseresponsestateset}{\noiseresponsestate(1,k)}
\newcommand*{\allresponsestates}{\Omega_{\system}}

\newcommand*{\connectedness}{\mathpzc{c}}
\newcommand*{\triggering}{\mathpzc{t}}

\newcommand*{\perception}{\mathpzc{p}}
\newcommand*{\Perception}{\mathcal{P}}
\newcommand*{\differentiation}{\mathcal{D}}
\newcommand*{\perceptiondifferentiation}{\differentiation_{\perception}}
\newcommand*{\evokeddifferentiationcapacity}{\overline{\differentiation}}
\newcommand*{\evokedperceptiondifferentiationcapacity}{\overline{\perceptiondifferentiation}}
\newcommand*{\intrinsicdifferentiationcapacity}{\differentiation^{\mathrm{max}}}
\newcommand*{\differentiationstructure}{C_{\differentiation}}

\newcommand*{\Matching}{\mathcal{M}}

\newcommand*{\stimvar}{\sensoryinterface_{t-\tau} = \stimulus}
\newcommand*{\mechvar}{M_t = m}
\newcommand*{\conditional}{\Pr(\mechvar\mid\stimvar)}
\newcommand*{\marginal}{\Pr(\mechvar)}

\newcommand{\autocite}{\cite}
\newcommand{\textcite}{\cite}
\newcommand*{\fullref}[1]{\hyperref[{#1}]{\cref*{#1},~\nameref*{#1}}}
\newcommand*{\fullrefnocomma}[1]{\hyperref[{#1}]{\cref*{#1}~\nameref*{#1}}}

\newcommand*{\argmax}{\operatornamewithlimits{argmax}\limits}

\usepackage[most]{tcolorbox}

\usepackage{subcaption}

\begin{document}
\vspace*{0.2in}

\begin{flushleft}
{\Large\textbf\newline{
    Intrinsic meaning, perception, and matching
}}\newline
\\
William G.P.\ Mayner\textsuperscript{1},
Bjørn Erik Juel\textsuperscript{1,2,3},
Giulio Tononi\textsuperscript{1,}*
\\
\bigskip
{\footnotesize
\textbf{1} Wisconsin Institute for Sleep and Consciousness, University of Wisconsin--Madison, Madison, WI, USA
\\
\textbf{2} Brain Signaling Group, University of Oslo, Oslo, Norway
\\
\textbf{3} Vestre Viken Kognitiv Nevrovitenskap, Vestre Viken Health Trust, Drammen, Norway
\\
\bigskip

%
%

* gtononi@wisc.edu
}

\end{flushleft}

\section*{Abstract}
\label{matching:secabstract}

Integrated information theory (IIT) argues that the substrate of consciousness is a maximally irreducible complex of units. Together, subsets of the complex specify a cause--effect structure, composed of distinctions and their relations, which accounts in full for the quality of experience. The feeling of a specific experience is also its meaning for the subject, which is thus defined intrinsically, regardless of whether the experience occurs in a dream or is triggered by processes in the environment. Here we extend IIT's framework to characterize the relationship between intrinsic meaning, extrinsic stimuli, and causal processes in the environment, illustrated using a simple model of a sensory hierarchy. We argue that perception should be considered as a structured interpretation, where a stimulus from the environment acts merely as a trigger for the complex's state and the structure is provided by the complex's intrinsic connectivity. We also propose that perceptual differentiation---the richness and diversity of structures triggered by representative sequences of stimuli---quantifies the meaningfulness of different environments to a complex. In adaptive systems, this reflects the “matching” between intrinsic meanings and causal processes in an environment.

\section*{Author summary}
\label{matching:secauthor-summary}

Here, we extend the integrated information theory of consciousness to characterize how intrinsic meanings are triggered by extrinsic stimuli. Using simple simulated systems, we argue that perception is a structured interpretation of a system state, where the state is triggered by a stimulus, but the interpretation is provided by a system’s intrinsic connectivity. We then propose that the “matching'' between a system and an environment can be measured by assessing the richness and diversity of intrinsic meanings triggered by representative sequences of stimuli. This approach offers a way of understanding how the meaning of an experience, which is necessarily intrinsic to the subject, can refer to extrinsic entities or processes.


\section{Introduction}
\label{matching:sec:introduction}

Cast a glance at the scene outside the window. In a blink of the eye, you see the forest with its intricate canopy of trees. How does this come about? A standard account is that a stimulus from the environment impinges on the retina, conveying information to the brain; the information is processed through a hierarchy of sensory areas, aided by top-down signals that try to predict, fill in, or disambiguate noisy bottom-up data; and finally, the meaning of the information is decoded, with the ultimate goal of guiding behavior. The very idea of processing suggests that the information is in the stimulus, ready to be decoded, perhaps with the help of ``contextual'' information from prior knowledge, and that meaning is in an activity pattern or ``code'' resulting from that processing. Somewhere along this processing chain, some of this information happens to become conscious (``conscious processing'').

Integrated information theory (IIT; \textcite{albantakis2023integratedb}) offers a different account (for a comparison with information processing views, see \autocite{zaeemzadeh2024shannon}). IIT starts from an experience---whether dreamt, imagined, or triggered by a stimulus---which it characterizes as a \emph{cause-effect structure}, or  \emph{\phistructure{}}. This is supported by a maximally irreducible substrate, or \emph{complex}, in a given state. The \phistructure{} is composed by distinctions and relations that define the feeling of the experience in a way that is fully intrinsic, without any reference to anything outside the complex \autocite{tononi2022only, tononiforthcomingbeing}. From the intrinsic perspective of an experiencing subject, the feeling of the experience is also its intrinsic meaning: What any content of the experience feels like---a distant sound, an imagined triangle, or a sense of nausea---is also what that content means for the subject. Thus, IIT views external stimuli as triggers of \emph{intrinsic meaning}, rather than as sources of information to be processed \cite{zaeemzadeh2024shannon}.

What, then, is the relationship between experiences and the stimuli triggering them? And how does the intrinsic meaning of experiences reflect features of the environment? In this paper, we extend IIT’s framework to address these questions. We present the mathematical formalism in the \nameref{matching:sec:theory} section. In the \nameref{matching:sec:results} section, we demonstrate the formalism with simple model systems. First, we briefly summarize how a complex in a given state, disconnected from the environment, supports a \phistructure{} that corresponds to a ``dreaming'' experience, which fully specifies its intrinsic feeling/meaning \autocite{ellia2021consciousness}. We then connect the complex to the environment, present a stimulus through the sensory interface, and let the complex quickly settle into its ensuing state. Using IIT’s actual causation formalism \autocite{albantakis2019what}, we calculate the \emph{triggering coefficient}---the extent to which the current state of each subset of the complex is caused by the stimulus. We can then unfold the \phistructure{} specified by the triggered state and obtain the \textit{perceptual structure}---the portion of the \phistructure{} triggered by the stimulus---and its  \emph{perceptual richness}---the quantity of intrinsic feeling/meaning triggered. Perception should then be considered as a structured interpretation of a state of the complex triggered by a stimulus---a structure whose feeling/meaning is determined intrinsically by the connectivity of the complex. In a well-adapted system, whose intrinsic connectivity was molded by evolution, development, and learning, that interpretation can be expected to “represent” relevant causal features of the environment. In general, however, the representation will not be straightforward. Finally, by exposing a complex to a sequence of stimuli representative of an environment, we calculate \emph{perceptual differentiation} as the perceptual richness of the union of the triggered perceptual structures. By capturing the richness and diversity of intrinsic meanings triggered by various features of a given environment, perceptual differentiation quantifies the meaningfulness of different stimulus sequences to the subject of experience. In a well-adapted brain, perceptual differentiation will reflect the \emph{matching} between intrinsic meanings and causal features of an environment.

\section{Theory}
\label{matching:sec:theory}

\subsection{The system and its environment}
\label{matching:sec:theory:overview}

In IIT, physical existence is synonymous with having cause--effect power, the ability to take and make a difference. A physical substrate $U$ with state space $\Omega_U$ is operationally defined by its potential causal interactions, assessed in terms of conditional probabilities. Accordingly, as in our prior work \autocite{albantakis2023integratedb}, the starting point of our analysis is a stochastic system
\(
U = \{ U_1, U_2, \ldots, U_n \}
\)
of $n$ discrete interacting binary units with state space \(
\Omega_U = \prod_i \Omega_{U_i}
\)
and current state $u \in \Omega_U$. We denote the complete transition probability function of $U$ over a system update $u \rightarrow \ubar$ as
\begin{align}
    \tpm{U} \;\equiv\; \Pr(\ubar \given u), \quad \forall \, u, \ubar \in \Omega_U.
\end{align}
We assume that the system state updates in discrete steps, that
the state space $\Omega_U$ is finite, and that the individual random variables
$U_i \in U$ are conditionally independent from each other given the preceding
state of $U$:
\begin{align}
\label{eq:conditional-independence}
    \Pr(\overline{u} \given u) \;=\; \prod^{n}_{i=1} \,\Pr(\overline{u}_i \given u).
\end{align}
We also assume a complete description of the system, meaning that we can determine the conditional probabilities in \eqref{eq:conditional-independence} for every state, with $\Pr(\bar{u} \given u) = \Pr(\bar{u} \given \mathrm{do}(u))$, where the ``do-operator'' indicates that $u$ is imposed by intervention \autocite{albantakis2019what, ay2008information, pearl2009causality, janzing2013quantifying}. This implies that $U$ is a causal network \autocite{albantakis2019what} and $\tpm{U}$ is a transition probability matrix (TPM) of size $\card{\Omega_U}$.

In this work, we divide $U$ into two parts: the system in
question $S \subseteq U$ and its environment $\environment = U \setminus S$. We define the \emph{sensory interface}
$\sensoryinterface \subseteq \environment$ to be the part of the environment that has an effect
on $S$ over one update step, such that the next state of $S$ depends only on its own
current state and the current state of $\sensoryinterface$. Throughout, we use \emph{stimulus} to refer to the state of the sensory interface.

\subsection{Experience and intrinsic meaning}
\label{matching:sec:theory:experience}

In this section we briefly recapitulate IIT's account of consciousness. For a complete description, we refer the reader to \autocite{albantakis2023integratedb, hendren2024iit}.

IIT identifies five essential properties of consciousness (\textit{intrinsicality}, \textit{information}, \textit{integration}, \textit{exclusion}, and \textit{composition}) that are immediately given and irrefutably true of every conceivable experience, termed `axioms': phenomenal experience is
\begin{enumerate*}[label=(\arabic*)]
    \item \emph{intrinsic} (it exists \emph{for itself})
    \item \emph{specific} (it is \emph{this one})
    \item \emph{unitary} (it is \emph{a whole}, irreducible to its parts)
    \item \emph{definite} (it is \emph{this whole}, containing all it contains, neither less nor more)
    \item \emph{structured} (it is composed of \emph{distinctions} and the \emph{relations} that bind them together, yielding a \emph{phenomenal structure} that feels \emph{the way it feels})
\end{enumerate*}.
The theory then aims to account for these phenomenal properties in ``physical'' terms, understood operationally, by postulating that the substrate of experience must jointly possess certain causal properties that correspond to them. IIT formulates these properties as its five `postulates' (\emph{intrinsicality}, \emph{information}, \emph{integration}, \emph{exclusion}, and \emph{composition}). The postulates take mathematical form, and IIT provides an algorithm for operationally assessing the extent to which a candidate substrate satisfies them \cite{mayner2018pyphi}. According to IIT, the physical properties characterized by the postulates are necessary and sufficient for a system to be a substrate of consciousness. Furthermore, IIT proposes an  explanatory identity: the quality of an experience can be accounted for in full  by the \phistructure{} unfolded from a \textit{maximal substrate}, or \emph{complex} (defined below). According to this identity, all phenomenal properties of experience must have a good explanation in terms of properties of the corresponding \phistructure{}, with no additional ingredients required \autocite{haun2019why, comolatti2024why}.

\subsubsection{Identifying the substrate of consciousness}
\label{matching:sec:theory:identifying-the-psc}

The first step in the analysis is to identify the substrate of consciousness. This is done by applying the first four postulates. The \textbf{intrinsicality} postulate requires that a system exert cause--effect power \emph{within itself}. To assess this, we causally marginalize the background units of the system $W = U \setminus S$, conditional on their current state, which renders them causally inert with respect to $S$.  The \textbf{information} postulate requires that a system's cause--effect power be specific: the system in its current state $s$ must select a specific cause--effect state for its units. This is the state for which \emph{system intrinsic information} ($\mathrm{ii}_s$) is maximal \autocite{barbosa2020measure}. The \textbf{integration} postulate requires that the system specify its cause--effect state in a way that is irreducible---\ie{}, that cannot be reduced to the joint specification of its parts. This is assessed by \emph{partitioning} the system into parts and quantifying how much system intrinsic information is lost due to the partition; \emph{system integrated information} ($\phi_s$) is then evaluated as the amount lost over the partition that makes the least difference \autocite{marshall2023system}. Finally, the \textbf{exclusion} postulate requires that the substrate of consciousness be constituted of a definite set of units, all of them, neither less nor more. Moreover, the units and updates of the substrate must have a definite grain. Since multiple overlapping subsets of $U$ may have a positive value of $\phi_s$, exclusion is enforced operationally by selecting the set of units that maximizes $\phi_s$ over itself. This set of units is called a \emph{maximal substrate} or \emph{complex}.

Again, we refer the reader to \textcite{albantakis2023integratedb} for definitions of these quantities and a full description of the formalism.

\subsubsection{Unfolding the \phistructure{} and characterizing intrinsic meaning}
\label{matching:sec:theory:unfolding-the-phi-structure}

Once a complex has been identified, we apply the postulate of \textbf{composition}. This requires that we characterize the complex's \emph{cause--effect structure}, or \emph{\phistructure{}}, \footnote{We use ``cause--effect structure'' to refer to the unfolded cause--effect power of a given system, whether or not it has been identified as a complex. If the system is a complex, we may use the term ``\phistructure{}.''} by considering all its subsets and \emph{unfolding} its cause--effect power. We denote the \phistructure{} of a complex $S$ in state $\responsestate$ as $\ces(\responsestate)$.

To contribute to the \phistructure{} of a complex, a system subset must both take and make a difference within the system. A subset $M \subseteq S$ in state $m \in \Omega_M$ is called a \emph{mechanism} if it \emph{links} a cause and effect state over subsets of units $Z_\mathrm{c} \subseteq S$ and $Z_\mathrm{e} \subseteq S$, called the cause and effect \emph{purviews}. A mechanism together with its cause and effect is called a \emph{causal distinction}, denoted $d(m)$. Distinctions are again evaluated based on whether they satisfy the postulates of IIT (except for composition, as distinction are themselves components). Briefly, for each candidate purview $Z_{c/e} \subseteq S$, the cause and effect states $z_c$ and $z_e$ are those for which the mechanism specifies maximal intrinsic cause information ($\mathrm{ii}_c$) and effect information ($\mathrm{ii}_e$), respectively \autocite{barbosa2020measure}. Integration is then assessed for a candidate purview by evaluating all partitions of the mechanism and purview, where the associated integrated information $\phi_{c/e}$ is the amount of $\mathrm{ii}_{c/e}$ lost over the partition that makes the least difference. Next, exclusion is enforced by selecting from among the candidate purviews $Z_{c/e}$ the cause and effect purview that respectively maximize the $\phi_c$ and $\phi_e$ values, yielding the cause purview state $z^*_c(m)$ and effect purview state $z^*_e(m)$. The integrated information of the distinction is the minimum of the cause and effect integrated information, $\phi_d(m) = \min(\phi_c(m), \phi_e(m))$, which quantifies its irreducibility. Finally, by the information postulate, the distinctions that exist for the complex are only those whose cause--effect state is congruent with that of the complex as a whole.

A set of distinctions $\mathbf{d}$ are \emph{bound} together by a \emph{causal relation} $r(\mathbf{d})$ if the cause--effect state of each distinction $d \in \mathbf{d}$ overlaps congruently over a shared set of units, called the \emph{relation purview}, which may be part of the cause, effect, or both the cause and effect of each distinction.
For a given set of distinctions, there are potentially many ``relating'' sets of causes and/or effects $\mathbf{z}$ with non-empty intersection. These specify unique aspects about the relation $r(\mathbf{d})$ and constitute its \emph{faces} $\mathbf{f}(\mathbf{d})$ (by analogy to the faces of a simplex). For a given face $f \in \mathbf{f}(\mathbf{d})$ the set $\mathbf{z}$ is called the \emph{face purview}; the union of the face purviews forms the \textit{relation purview}.
A relation $r(\mathbf{d})$ that binds together $h = \card{\mathbf{d}}$ distinctions is called an \emph{$h$-degree relation}, and a relation face $f(\mathbf{d}) \in \mathbf{f}(\mathbf{d})$ over $k = \card{\mathbf{z}}$ cause/effect purviews is called a \emph{$k$-degree face}.
Briefly, the irreducibility $\phi_r(\mathbf{d})$ of a causal relation is measured by ``unbinding'' distinctions from their joint purviews, taking into account all faces of the relation, as follows. For each distinction $d \in \mathbf{d}$, the average integrated information $\varphi_d$ per unique distinction purview unit is multiplied by the number of units in the relation purview. The minimum of this value across distinctions is the relation's irreducibility $\phi_r(\mathbf{d})$, corresponding to the integrated information lost by partitioning that distinction from the relation.

The \phistructure{} of the complex in its current state is composed of these distinctions and relations. By the explanatory identity of IIT, these account in full, with no additional ingredients, for the quality (or feeling) of an experience, which is the same as its intrinsic meaning (``the meaning is the feeling''). The sum of integrated information values $\phi$ of all the distinctions and relations that compose the \phistructure{}, called \emph{structure integrated information}, denoted $\Phi$, corresponds to the quantity of consciousness.

\subsubsection{Distinction \phifolds{}}
\label{matching:sec:theory:distinction-phi-folds}

Here we introduce a convenient decomposition of the \phistructure{} $\ces(\responsestate)$ into sub-structures, called \emph{\phifolds{}}, associated with each distinction.

For a given mechanism $M \subseteq S$ in state $m$ that specifies a distinction $d(m)$, we call the sub-structure consisting of that distinction and all relations involving it the \emph{distinction \phifold} of $d(m)$, denoted $\ces(d(m))$, or simply $\ces(d)$ when $d$ ranges over distinctions. If the \phistructure{} is thought of as a hypergraph, with distinctions as vertices and relations as edges, then a distinction \phifold{} corresponds to a single vertex and its incident edges.

We can define a quantity $\Phi_d(\ces(d(m)))$ that captures the contribution of the mechanism $m$ to the total $\Phi(\ces(\responsestate))$ as
\begin{equation}
    \label{eq:distinction-phi-fold}
    \Phi_d(\ces(d(m))))
        \;=\; \sum_{c \,\in\, \ces(d(m))} \frac{
            \phi_c
        }{
            \card{c}
        },
\end{equation}
where $c$ denotes an arbitrary \emph{component} of the \phistructure{} (distinction or relation). For a distinction $c = d(m)$, $\card{c} = \card{d(m)} = \card{\{ m \}} = 1$. For a relation $c = r(\mathbf{d})$, $\card{c} = \card{r(\mathbf{d})} = \card{\mathbf{d}} \geq 1$; in words, the relation $\varphi$ value is divided by the number of distinctions it binds together.

This expression counts the entire $\phi_d$ value of the distinction, but a fraction of the $\phi_r$ of relations involving $d(m)$. This assumes that a relation's integrated information ($\phi_r$) is distributed uniformly across its distinctions. Since the relation is an irreducible component of the \phistructure{} and removing any one of the distinctions would ``unbind'' them, we consider the contribution of each $d \in \mathbf{d}$ to the relation's $\phi_r$ value to be $\phi_r / \card{\mathbf{d}}$.

We then have that the total $\Phi(\ces(\responsestate))$ is partitioned by the $\Phi_d$ values:
\begin{equation}
    \label{eq:distinction-phi-fold-partition}
    \Phi(\ces(\responsestate))
        \;=\; \sum_{d \,\in\, \ces(\responsestate)} \Phi_d(\ces(d)).
\end{equation}
In words, the $\Phi$ value of the entire \phistructure{} can be expressed as the sum of the $\Phi_d$ values of the distinction \phifolds{} specified by each of the system's irreducible mechanisms.

\subsection{Perception}
\label{matching:sec:theory:perception}

\subsubsection{Connectedness \& the triggering coefficient}
\label{matching:sec:theory:connectedness-and-triggering-coefficient}

Let $p$ be the conditional probability of $M = m$ at time $t$ given $\sensoryinterface = \stimulus$ at time $t - \tau$,
\[p = \conditional,\]
and let $q$ be the marginal probability of $M = m$ at time $t$,
\[q = \marginal,\]
where the marginalization is over possible stimuli $\Omega_{\sensoryinterface}$. Note that because we take the intrinsic perspective of the system under analysis, we assume the uniform distribution over stimuli, $\Pr(\stimulus) = 1/\card{\Omega_{\sensoryinterface}}$, rather than the observed distribution \autocite{albantakis2023integratedb}.

We define the \emph{connectedness} $\connectedness(\stimulus, m)$ of a set of units $M \subseteq S$ in state $m$ at time $t$ to the sensory interface $\sensoryinterface$ in state $\stimulus$ at $t - \tau$ as
\begin{align}
\label{eq:connectedness}
\connectedness(\stimulus, m)
    \;=&\;
    \begin{cases}
        \log_2 \left( p / q \right)  &\textrm{if } p > 0,\ q > 0, \textrm{ and } p \geq q, \\
        0                            &\textrm{otherwise}.
    \end{cases}
\end{align}
This definition is based on the measure of actual effect information developed in \textcite{albantakis2019what}, although here we do not consider the question of which subset of the sensory interface caused the current state of the subset (``what causes what'').\footnote{To exemplify, consider a subset $M$ of eight binary units such that a single stimulus $\sensoryinterface = x$ fully determines the state $M = m \in \{0, 1\}^8$ (so that $p = 1$), and $m$ never occurs in response to any other stimulus (so that $q = 1/2^{\card{\sensoryinterface}}$). If the sensory interface also consists of eight binary units, then $\connectedness(x, m) = 8$ bits. If $\card{\sensoryinterface} = 8$ but $\card{M} = 1$, we likewise have $\connectedness(x, m) = 8$ bits. But if $\card{\sensoryinterface} = 1$ and $\card{M} = 8$, then $\connectedness(x, m) = 1$ bit. In general, if $m$ always and only occurs in response to a single stimulus, connectedness is limited by the information specified by $\sensoryinterface = x$. Conversely, if $p = 1$ but $m$ occurs in response to other stimuli, then $q > 1/2^{\card{\sensoryinterface}}$, and the connectedness value is instead limited by the information specified by $M = m$.}
The expression in the first case is also known as the pointwise mutual information (PMI), and is positive when the stimulus raises the probability of the mechanism's state occurring compared to when the influence of the stimulus is ignored.\footnote{If $p < q$, the PMI is negative. In this case, we take the stance that the environment did not bring about the state $M = m$ and define connectedness to be zero, in line with the actual causation framework \autocite{albantakis2019what}, rather than considering this a ``preventative effect'' on the subset state \autocite{korb2011new}. In information-theoretic contexts, this measure has been termed the positive PMI \autocite{dagan1993contextual}.} It is important to note that because we use the uniform distribution as the prior $Pr(x)$, this is a causal, rather than a correlational, measure (in contrast to the PMI, as typically employed); hence our choice of the term ``connectedness''.

Connectedness $\connectedness(\stimulus, m)$ is maximized when $p = 1$, \ie, when the stimulus $\stimulus$ causes the mechanism state $m$ deterministically. It is therefore bounded by the self-information of the mechanism state:
\begin{equation}
    \label{eq:connectedness-max}
    \connectedness(\stimulus, m) \;\leq\; \log_2( 1 / q).
\end{equation}
We use this bound to define a normalized form of connectedness, called the \emph{triggering coefficient}:
\begin{equation}
\label{eq:triggering-coefficient}
    \triggering(\stimulus, m)
        \;=\;
        \frac{
            \connectedness(\stimulus, m)
        }{
            \log_2(1/q)
        }\;,
\end{equation}
so that $0 \leq \triggering(x, m) \leq 1$. When the mechanism $m$ specifies a distinction $d$, we will also refer to its triggering coefficient as $\triggering(x, d)$ or $\triggering(x, d(m))$.

The triggering coefficient expresses the extent to which the stimulus $\stimvar$ caused $\mechvar$ relative to how strong that cause could have been, taking into account the size of the mechanism as well as the system's intrinsic connectivity. This is because, while the connectedness value is dependent on the size of the subsets (larger subsets can specify more information), the triggering coefficient is independent of mechanism size; a value of $\triggering(\stimulus, m) = 1$ indicates that $\stimvar$ was fully sufficient to cause $\mechvar$, irrespective of the amount of information specified by $\mechvar$, while $\triggering(\stimulus, m) = 0$ indicates the stimulus had no role in bringing about $\mechvar$.\footnote{
There are other possible normalizations for the connectedness value. The positive PMI is also bounded by (1) the self-information of the stimulus, $-\log_2(\Pr(\stimulus))$; (2) the minimum of both self-information values, $\min(-\log(\Pr(\stimulus)), -\log(\Pr(m)))$; and (3) the joint self-information, $-\log_2(\Pr(\stimulus, m))$. Normalizing by each of these bounds respectively yields a measure of (1) non-degeneracy, \ie{} the selectivity of the mechanism in responding to the stimulus (maximized when $\Pr(\stimulus \given m) = 1$); (2) either determinism or non-degeneracy (maximized when either $\Pr(\stimulus \given m) = 1$ or $\Pr(m \given \stimulus) = 1$); and (3) determinism and non-degeneracy, \ie{} perfect co-occurrence (maximized when both $\Pr(\stimulus \given m) = Pr(m \given \stimulus) = 1$). Bouma (2009) noted these options and investigated the latter in the context of linguistics, calling it the normalized PMI \autocite{bouma2009normalized}. In our case, by contrast, the measure should be sensitive to determinism but not to degeneracy (when $\mechvar$ also occurs in response to other stimuli, raising the marginal probability $q$), so that, given a stimulus and a response that actually occurred, the triggering coefficient quantifies the causal role of the stimulus in producing the response regardless of whether other stimuli might have also caused the response counterfactually.
}

The triggering coefficient also permits the unbiased comparison of different subsets by accounting not only for subset size,\footnote{This is crucial because, for IIT, what matters is only whether a subset has irreducible cause--effect power within the system, not how many units it contains.} but also for the role of the system's intrinsic connectivity. In general, subsets that are more directly connected to the sensory interface will be more strongly connected to the environment, while internal subsets will tend to be affected more by intrinsic dynamics. Normalizing by the self-information of the mechanism state, $\log_2(1/q))$, discounts such differences. For example, a neuronal assembly deep in the brain whose state is determined largely by intrinsic dynamics (\ie, connectedness is low in absolute terms), but which is nonetheless reliably triggered by a particular stimulus (connectedness is high relative to other stimuli), will be evaluated on the same terms as an assembly in earlier sensory areas.

Note that $\tau$, the delay at which the stimulus's effect is evaluated for all subsets, is a free parameter and should be chosen to maximize the efficacy of the stimulus. In practice, the appropriate choice of $\tau$ will depend on the experimental setup.

\subsubsection{Perception}
\label{matching:sec:theory:perception-and-perceptual-richness}

We define the \emph{perception value} of a distinction $d(m)$ as
\begin{equation}
    \label{eq:perception-distinction}
    \perception(\stimulus, d(m))
        \;=\; \triggering(\stimulus, m) \; \phi_d(m).
\end{equation}
In words, a distinction with $\phi_d(m) > 0$ can contribute to the perception value to the extent that it was triggered by the stimulus. The perception value is zero ($\perception(\stimulus, d(m)) = 0$) when either the subset of the system $m$ is reducible and does not specify a distinction ($\phi_d(m) = 0$), or when the stimulus did not have an effect on the subset ($\triggering(\stimulus, m) = 0$). Note that because $\triggering(\stimulus, m) \leq 1$, the maximum perception value for a distinction is its $\phi_d$ value.

This approach can be straightforwardly extended to the case of relations. Since each relation $r(\mathbf{d})$ binds together several distinctions $\mathbf{d}$, we can define the triggering coefficient for a relation $\triggering(\stimulus, r(\mathbf{d}))$ as a weighted average of the triggering coefficients of each $d \in \mathbf{d}$. To determine the weights, we use the same rationale as for the definition of the $\Phi_d$ of a distinction \phifold{}: since the relation is irreducible, we assume that its integrated information $\phi_r(\mathbf{d})$ is distributed uniformly across the distinctions it binds together (as removing any one of them would ``unbind'' all of them). Thus, each mechanism's triggering coefficient is given equal weight in the average:
\begin{equation}
    \label{eq:triggering-relation}
    \triggering(\stimulus, r(\mathbf{d}))
        \;=\;
        \frac{1}{\card{\mathbf{d}}} \sum_{d \,\in\, \mathbf{d}} \triggering(\stimulus, d).
\end{equation}
The perception value of a relation $r(\mathbf{d})$ with respect to a stimulus $\stimulus$ is then defined as
\begin{equation}
    \label{eq:perception-relation}
    \perception(\stimulus, r(\mathbf{d}))
        \;=\;
        \triggering(\stimulus, r(\mathbf{d})) \; \phi_r(\mathbf{d}).
\end{equation}
Note that for distinction \phifolds{}, this implies
\begin{equation}
    \label{eq:perception-phi-fold}
    \sum_{c\,\in\,\ces(d(m))}\perception(\stimulus, c)
        \;=\;
        \triggering(\stimulus, m) \; \Phi_d(\ces(d(m))).
\end{equation}

\eqs \eqref{eq:perception-distinction} and \eqref{eq:perception-relation} can be combined into a general expression for an arbitrary component $c$ of the \phistructure{} (distinction or relation):
\begin{equation}
    \label{eq:perception-component}
    \perception(\stimulus, c)
        \;=\;
        \triggering(\stimulus, c)
        \frac{\phi_c}{\card{c}},
\end{equation}
where $\card{c}$ denotes the number of distinctions involved in the component, as in \eqref{eq:distinction-phi-fold}.

Next, we consider the relationship of the stimulus to the \phistructure{} as a whole. Analogously to (structure) integrated information $\Phi$---the value of integrated information for a \phistructure{}, we define  \emph{perceptual richness} $\Perception$ to be the sum of the perception values of the components in the specified \phistructure{}:
\begin{equation}
    \label{eq:perception-richness}
    \Perception(\stimulus, \responsestate)
        \;=\;
        \sum_{c \,\in\, \ces(\responsestate)} \perception(\stimulus, c).
\end{equation}
For a distinction \phifold{}, we denote the sum of perception values as $\Perception_d(\stimulus, m)$. We call the \phistructure{}, weighted by the triggering coefficients of all its components, a \textit{perceptual structure} (and relevant sub-structures, \textit{percepts}). Note that by \eqs \eqref{eq:distinction-phi-fold} and \eqref{eq:perception-component}, a perceptual structure can be partitioned into the $\Phi_d$ values of each distinction \phifold{} in $\ces(\responsestate)$, weighted by their triggering coefficients:
\begin{equation}
    \label{eq:perception-richness-phi-fold}
    \Perception(\stimulus, \responsestate)
        \;=\;
        \sum_{m \,\subseteq\, \responsestate} \Perception_d(\stimulus, m)
        \;=\;
        \sum_{d \,\in\, \ces(\responsestate)} \triggering(\stimulus, d) \,\Phi_d(\ces(d))
\end{equation}

Perceptual richness thus quantifies the extent to which an external stimulus triggered an experience within a complex in a state---which defines the intrinsic meaning for the complex\autocite{tononi2022only, tononiforthcomingbeing}.

\subsection{Perceptual differentiation}
\label{matching:sec:theory:differentiation}

We now extend our analysis to a sequence of $k$ stimuli $\stimulusset = \left( \stimulus_1, \stimulus_2, \ldots, \stimulus_i, \ldots, \stimulus_k \right)$ that triggers the sequence of system states $\responsevarset = \left( \responsevar_1, \responsevar_2, \ldots, \responsevar_i, \ldots, \responsevar_k \right)$.

First we introduce the notion of the \emph{differentiation} triggered by $\stimulusset$. The differentiation of a particular system response sequence $\responsevarset = \responsestateset$ triggered by $\stimulusset$ measures the diversity of the corresponding set of \phistructures{}.

For the purposes of this work, we define differentiation as follows.\footnote{The question of how to best measure the diversity of a set of \phistructures{} in general is an interesting and difficult one, which we leave for future work. Importantly, because differentiation is defined operationally for an observer, its definition is ultimately a matter of experimental methodology---so, unlike the quantities of IIT, which aim to account for consciousness itself and thus must be unique and are either correctly formulated or not, there need not be a unique and correct measure of differentiation. Accordingly, here we define a straightforward measure that suffices for illustrating the idea of \emph{matching} in our examples. We further note that for practical purposes, differentiation can be estimated using only the activity patterns triggered by the stimuli (\ie, without computing the corresponding \phistructures{}), as in \autocite{mayner2022measuring,mensen2017eeg,mensen2018differentiation,mensen2018differentiationa,marshall2016integrated,boly2015stimulus}; this is because the diversity of \phistructures{} will generally be highly (though not linearly) correlated to that of the activity patterns that determine them (modulo changes over time in the causal structure of the system).} We count every unique component $c$ that appears in the associated set of \phistructures{}; that is, we consider the union of the \phistructures{} associated with $\responsestateset$, called the \emph{differentiation structure}:
\begin{equation}
    \label{eq:differentiation-structure}
    \differentiationstructure\left(\responsestateset\right)
        \;=\;
        \bigcup_{i=1}^k\;
        \left\{c \mid c \,\in\, \ces(\responsestate_i)\right\}.
\end{equation}
The amount of \emph{differentiation} is then defined analogously to $\Phi$ as
\begin{equation}
    \label{eq:differentiation}
    \differentiation\left(\responsestateset\right)
        \;=\;
        \sum_{c\,\in\,\differentiationstructure\left(\responsestateset\right)} \phi_c.
\end{equation}

We can then define the \emph{extrinsic differentiation capacity} of a system as the total differentiation triggered by all possible stimuli:
\begin{equation}
    \label{eq:evoked-differentiation-capacity}
    \evokeddifferentiationcapacity(\system)
        \;=\;
        \differentiation(
            \{\responsestate \in \allresponsestates \given \exists \stimulus \in \Omega_{\sensoryinterface} \,\text{ such that }\, \Pr(\responsestate \given \stimulus) > 0\}
        )
\end{equation}
where $\allresponsestates$ denotes the set of all possible system states (note that order does not matter here).

We can also define the \emph{intrinsic differentiation capacity} of a system as the differentiation of all its possible states (including those that cannot be evoked by stimuli), which is the maximum differentiation the system can attain:
\begin{equation}
    \label{eq:intrinsic-differentiation-capacity}
    \intrinsicdifferentiationcapacity(\system)
        \;=\;
        \differentiation(\allresponsestates).
\end{equation}

We define the \textit{perceptual differentiation structure} similarly to the differentiation structure, and the associated \emph{perceptual differentiation} triggered by $\stimulusset$ similarly to differentiation, using perception values in place of $\phi$ values. Since for each occurrence of a given component $c \in \differentiationstructure\left(\responsestateset\right)$ the triggering coefficient will generally differ depending on the stimulus $x_i$, the perception value of a component is taken to be its maximum value across all $i$:
\begin{equation}
    \label{eq:perception-differentiation}
    \perceptiondifferentiation\left(\stimulusset, \responsestateset\right)
        \;=\;
        \sum_{c\,\in\,\differentiationstructure\left(\responsestateset\right)}
        \max_{i=1, \ldots, k}\;
        \perception\left(\stimulus_i, c\right),
\end{equation}
where $\perception(x_i, c) = 0$ if $c \notin C(y_i)$. The \emph{perceptual differentiation capacity} is then defined as
\begin{equation}
    \label{eq:evoked-perception-differentiation-capacity}
    \evokeddifferentiationcapacity(\sensoryinterface, \system)
        \;=\;
        \perceptiondifferentiation\Big(
            \left\{
                (\stimulus, \responsestate) \given \stimulus \in \Omega_{\sensoryinterface},\ \responsestate \in \Omega_\system \,\text{ such that }\, \Pr(\responsestate \given \stimulus) > 0
            \right\}
        \Big)
\end{equation}

By considering the union of the perceptual structures triggered by each stimulus, perceptual differentiation measures the extent to which the sequence triggers perceptual structures that are both rich and diverse.

\subsection{Matching}
\label{matching:sec:theory:matching}

Finally, we consider perceptual differentiation when a complex samples stimulus sequences from an environment through its sensory interfaces and actions. We assume that the environment (or ``world'') is much broader than the complex in terms of the number of units that constitute it and much deeper in terms of the length of the sequence of states it can visit.

We define the \emph{matching} between a complex and an environment as
\begin{align}
    \label{eq:stimulus-sequence-matching}
    \Matching\left(\sensoryinterface(1,k), S\right)
        \;=\;
        \max_{a, b \,\in\, (1,\,\ldots,\, k), \; a < b}
        \;\expectation
        \left[\,
            \perceptiondifferentiation\left(\sensoryinterface(a,b), \responsevar(a,b)\right)
            - \perceptiondifferentiation\left(\noisestimulus(a,b), \noiseresponsestate(a,b)\right)
        \,\right],
\end{align}
where $a$ and $b$ indicate the range of a contiguous subsequence of $(1, ..., k)$.

In words, matching is the maximum expected difference between the perceptual differentiation triggered in a complex by stimulus sequences sampled from an environment and that triggered by random stimulus sequences (``noise'' $N$).\footnote{
Note that the distribution of environmental stimuli is generally non-stationary.} It follows that, for matching to be high, environmental stimulus sequences must trigger more intrinsic meanings than would be expected by chance. This can only be so if the intrinsic meanings supported by the complex ``match'' or ``resonate'' with causal processes in the environment that generate stimulus sequences different from chance.

The reasoning for taking the maximum over contiguous subsequences is as follows. Since every possible stimulus has nonzero probability in $N$, as the sequence length $k$ grows, the probability that all possible stimuli occur in $N(1, k)$ approaches 1, so that
\begin{equation}
    \lim_{k \to \infty}
        \perceptiondifferentiation\left(\noisestimulusset, \noiseresponsestateset\right)
        =
        \evokedperceptiondifferentiationcapacity(\sensoryinterface, \system).
\end{equation}
That is, the differentiation evoked by random stimulus sequences approaches the system's evoked perceptual differentiation capacity. This implies that
\begin{equation}
    \label{eq:noise-limit}
    \lim_{k \to \infty}
    \expectation
    \left[\,
            \perceptiondifferentiation\left(\sensoryinterface(1, k), \responsevarset\right)
            - \perceptiondifferentiation\left(\noisestimulusset, \noiseresponsestateset\right)
    \,\right]
    \leq 0.
\end{equation}
(If every possible stimulus sequence also has nonzero probability in $E$, then \eqref{eq:noise-limit} becomes an equality.) Therefore, for completeness, the maximum is formally taken over contiguous subsequences so that the measure behaves as intended even for large $k$.
However, since the probability of a given stimulus being sampled from $N$ decreases exponentially as $\card{\sensoryinterface}$ increases, this limit is approached very slowly; for sequence lengths of any practical relevance in realistic systems, the maximum will likely be attained with the full sequence.

In practice, any stimulus sequences used for assessing perceptual differentiation experimentally in real systems will be relatively short, putting a premium on choosing a ``representative'' sampling based on heuristic criteria. A high value of perceptual differentiation and rapid growth with the length of stimulus sequences $k$ can provide a useful index of the representativeness of the sampling.

In general, for matching to be high, the following conditions must hold: \begin{enumerate*}[itemjoin={;\ }, itemjoin*={,\ and\ }]
    \item the system is a large complex, \ie{} one that can specify rich and diverse \phistructures{}
    \item the sampled environment contains rich and diverse causal processes that constrain the stimulus sequences impinging on sensory interfaces
    \item the system is strongly connected to the environment across most of its subsets
    \item the system has adapted its internal organization such that stimulus sequences sampled from its environment trigger richer and more diverse perceptual structures that those triggered by other environments or than would be expected by chance
\end{enumerate*}.

\section{Results}
\label{matching:sec:results}

As a proof of principle, we apply the formalism introduced above in two simple \emph{in silico} model systems. First, we apply the IIT analysis \autocite{albantakis2023integratedb} to unfold the system's intrinsic cause--effect structure (or \phistructure{}; for the present purposes, we treat the systems as parts of larger complexes, without evaluating maximal irreducibility for all states), which fully characterizes the intrinsic meaning specified by the system in its current state. Next, we demonstrate how our formalism characterizes perception by measuring the extent to which parts of the system's \phistructure{} are triggered by stimuli sampled from a simulated environment. We then demonstrate how, in our account, perception is best understood as interpretation, and we illustrate some implications of the formalism for notions of reference and representation. Finally, we demonstrate how the quantities \emph{perceptual differentiation} and \emph{matching} measure the extent to which an environment triggers rich and diverse percepts in a system.

\subsection{The substrate model}

\begin{figure}[H]
    \begin{adjustwidth}{-2.25in}{0in} 
    \centering
    \includegraphics[width=\maxwidth]{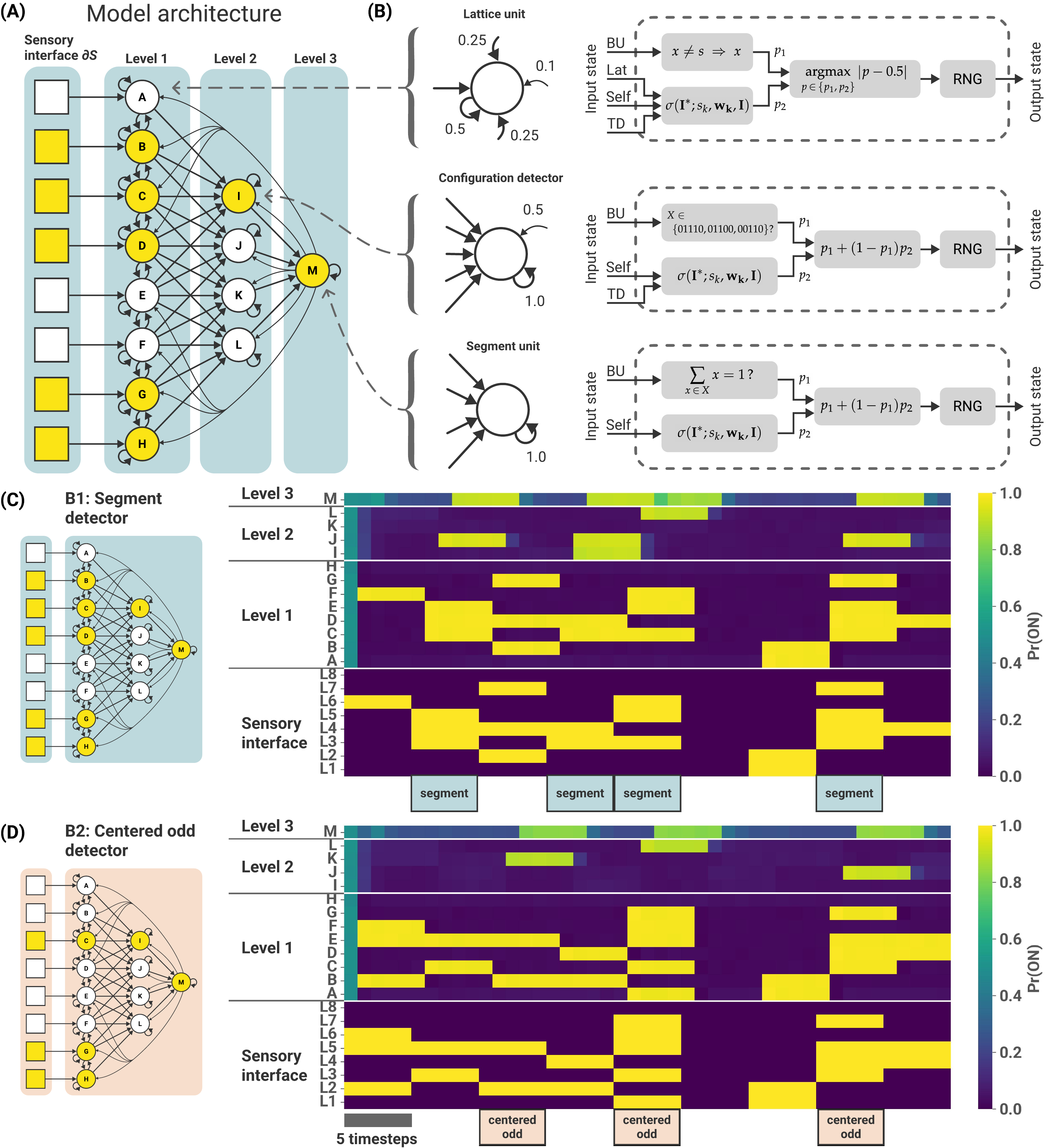}
    \caption[]{
        \label{matching:fig:model}
        \textbf{The substrate models and their dynamics.}
        \textbf{(A)} Diagram of the model architecture, shared by the two systems, termed B1 and B2. The systems each consist of 13 stochastic binary units (labeled circles). Yellow indicates the `ON' state (`1'), white indicates `OFF' (`0'). Units are arranged in hierarchical levels in a rough analogy to the visual system (see main text for description).
        \textbf{(B)} Schematic depiction of the units at each level and their activation functions. Bottom-up connections (BU) are shown as straight arrows pointing right, lateral connections (Lat) are shown as curved arrows pointing up and down, the self-connection is shown as a looping arrow, and top-down connections (TD) from the segment unit are shown as curved arrows pointing left.
    }
    \end{adjustwidth}
\end{figure}

The substrate $U$ models a system $\system$ connected to an environment $\environment$ via a sensory interface $\sensoryinterface \subseteq \environment$ (\cref{matching:fig:model}A).

We designed two model systems, `B1' and `B2' (\cref{matching:fig:model}), that, by construction, detect different stimulus features. In what follows, we briefly describe the model systems and their units (for further details, see \nameref{matching:appendix}).

Both systems are simple hierarchical neural networks that share a common architecture: each comprises 13 units connected in a hierarchical, convergent pattern such that the higher-level units are sensitive to a specific feature of environmental stimuli. This architecture is loosely inspired by models of the mammalian visual system, with feed-forward connectivity converging on a single `concept cell'.
The system constituents are modeled as binary units---neurons that can be `ON' (\ie{}, fire action potentials) or `OFF'---whose activation functions are state-dependent (modeling a form of short-term plasticity; \nameref{matching:appendix}).
We chose to use models simple enough that their dynamics can be fully characterized by a TPM of tractable size, so as to permit exact computation of the quantities in the formalism.

The first system, `B1', was designed to detect the presence of a `segment' feature on the sensory interface: a pattern of 5 units in which 2 or 3 contiguous central units are ON and surrounded by OFF units (\ie, $01110$, $00110$, or $01100$). The second system, `B2', detects a `centered odd' stimulus feature: a pattern of 5 units in which an odd number of central units are ON and surrounded by OFF units (\ie, $01110$ or $00100$).

\begin{figure}[H]
    \begin{adjustwidth}{-2.25in}{0in} 
    \caption*{
        Numbers on lateral and top-down connections indicate the connection strength used in the sigmoidal module $\sigma$ of each activation function.
        To the right of the unit diagrams, activation functions are represented schematically as compositions of simpler functions.
        B1 and B2 differ only in the bottom-up input configurations that the L2 configuration detectors are selective for: in B1, these are 01110, 01100, and 00110; in B2, these are 01110 and 00100. For details, see \nameref{matching:appendix}. RNG: random number generator.
        \textbf{(C)} Example of each system's dynamics. The plot shows the average response for each unit across 5000 simulations in which the same arbitrary sequence of stimuli was presented to B1. Throughout this work, stimuli are presented at a macro-timescale $\tau$ by clamping the sensory interface state for $\tau = 5$ elementary timesteps, in order to allow the effects to percolate through the hierarchy (fixing the state is not strictly necessary, but was done to reduce the repertoire of possible stimuli from $(2^8)^5$ to $2^8$ for computational simplicity). Four of the stimuli presented here contained `segments' (highlighted in the bottom row). As activity percolates upwards, the level 1 lattice units are strongly driven by the bottom-up inputs and reproduce the input pattern with high probability, level 2 configuration detectors activate in the presence of a segment in their receptive field, and the level 3 segment unit \emph{M} preferentially activates in the presence of a segment anywhere on the sensory interface.
        \textbf{(D)} Same as (C), but for B2. A different arbitrary sequence of stimuli was presented, each lasting $\tau = 5$ timesteps. Here we see that this system's detector units and invariant unit activate in response to `centered odd' stimuli rather than `segments'.
    }
    \end{adjustwidth}
\end{figure}

Level 1 of the hierarchical architecture comprises `lattice' units that each receive (1) a bottom-up input from a distinct unit in the sensory interface; (2) lateral input from adjacent lattice units; and (3) top-down input from the top level of the hierarchy. These units---the system's 'ports-in'---are reliably activated by the bottom-up sensory input and remain active as long as the input is `ON'.

Level 2 of the hierarchy is composed of `configuration detectors,' each of which receives bottom-up input from five lattice units below it as well as top-down input from the top level unit. The configuration detectors are reliably activated when the five level 1 units in their receptive field (RF) form the pattern they are tuned to. The configuration detectors of B1 are tuned to `segments' ($01110$, $00110$, or $01100$); B2's configuration detectors are tuned to `centered odds' ($01110$ or $00100$).

Level 3 consists of a single `invariant' unit that receives bottom-up input from the four configuration detectors. It is reliably activated whenever a configuration detector in level 2 is active.

Thus, the state of the neural network is set by its inputs such that B1's overall role can be understood as detecting the presence of a `segment' feature anywhere on the sensory interface, while B2's function is to detect the presence of a `centered odd' feature anywhere on the sensory interface. \cref{matching:fig:model}A shows a diagram of the system; the activation functions of each class of units are shown schematically in \cref{matching:fig:model}B. Every unit also features a self-connection that tends to stabilize the unit's state.

In our simulations, a stimulus $x$ from the environment is presented to the system by clamping the sensory interface to the state $\sensoryinterface = x$ for $\tau = 5$ time steps and allowing the effects to percolate through the system.\footnote{In general, the sensory interface state need not be fixed for this duration. We used a fixed state for the duration of the delay $\tau$ in order to minimize the number of possible stimuli under consideration and thereby simplify the computations. For a given stimulus $x$, we denote this presentation procedure by $\sensoryinterface_{t-\tau} = x$.} While the system state stabilizes as the stimulus is held constant, the strength of lateral, top-down, and self connections is modulated dynamically in a state-dependent manner. This is in line with empirical data indicating that activity patterns triggered by sensory stimuli are typically amplified, rather than overridden, by lateral or top-down signals \autocite{chance1999complex, douglas1995recurrent, li2013intracortical, lien2013tuned, peron2020recurrent, pattadkal2022primate}. In the model, what is adjusted dynamically is not the activity level (the units are binary) but rather the sign and net strength of the interactions. These are set such that the interactions always \textit{endorse} the unit's current state (\nameref{matching:appendix}).
For example, if a unit's lateral input is `ON' and the unit itself is `ON', the net effect of the lateral input is excitatory. If a unit's input is `ON' but the unit itself is `OFF', then net effect of the lateral input becomes inhibitory. Furthermore, the net strength of the interaction differs for the four combinations of input state and unit state, reaching the highest strength when both the input and the unit are `ON'. In the brain, such a short-term change in the efficacy of interactions might be mediated by the opening of NMDA receptors in cortical synapses \autocite{mayer1984voltagedependent, nowak1984magnesium} (and see, for example, \autocite{sabatini2002life,nevian2004single}), or by the coupling of apical inputs and somatic activation \autocite{larkum1999new, larkum2004topdown}.
The strong positive modulation of the interactions among connected, co-active units reflects the importance of transmitting to downstream areas `ON' states rather than `OFF' states, given that the brain employs neuronal firing---which is energetically expensive---to signal important occurrences \autocite{balduzzi2013what}. It also reflects the fact that in sensory areas, connections are strongest among neighboring units, internalizing the smoothness of sensory inputs \autocite{field1987relations, hubel1962receptive}, in line with the principle that ``what fires together wires together'' \autocite{hebb2002organization}.\footnote{A noise input is also likely to reduce (structure) perception values by triggering distinctions that specify causes and effects that are incongruent with the entire system's cause--effect state \autocite{albantakis2023integratedb}.}

Examples of each system's dynamics when presented with stimuli are shown in \cref{matching:fig:model}C. The level 1 units reliably relay the stimuli from the sensory interface to the level 2 configuration detectors, which preferentially activate when they receive their preferred input configuration. The level 3 unit then activates when one of the configuration detectors is active, indicating the presence on the sensory interface of a `segment' (for B1) or a `centered odd' configuration (for B2).

\subsection{Every experience is intrinsic and structured: its meaning is its feeling}

IIT proposes an `explanatory identity' between experiences and \phistructures{}: the phenomenal qualities of an experience can be fully accounted for by the distinctions and relations composing the \phistructure{} \autocite{ellia2021consciousness}. This correspondence can be summed up by the maxim \emph{``quality is structure''}.
A corollary of this correspondence is that the meaning of contents of experience, including those triggered by an external stimulus, must also be accounted for by the \phistructure{}, rather than somehow attached to or derived from its external referent. Put succinctly, \emph{the meaning is the feeling}\autocite{tononi2022only, tononiforthcomingbeing}.

\begin{figure}[H]
\begin{adjustwidth}{-2.25in}{0in} 
\centering
\includegraphics[width=\maxwidth]{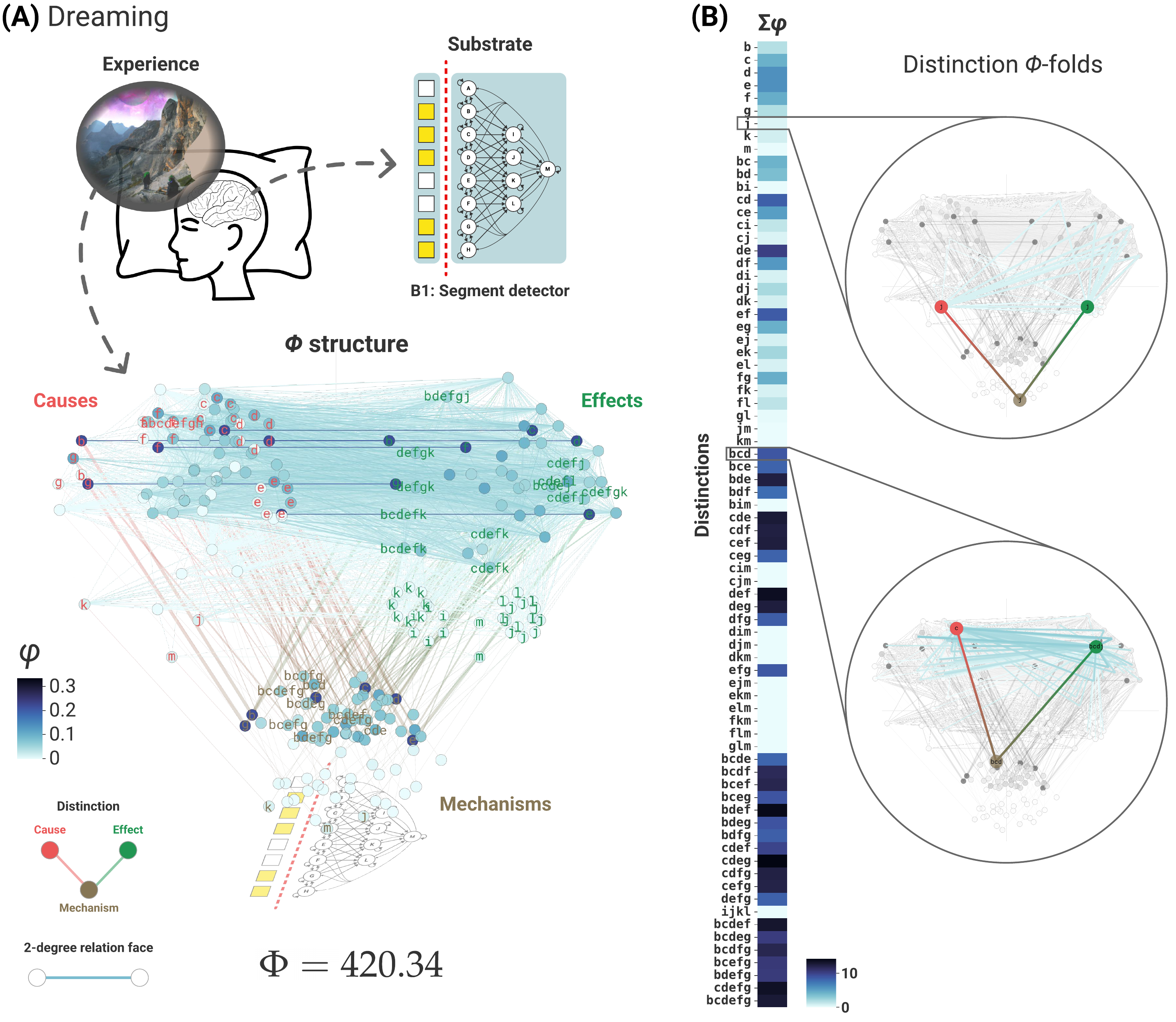}
\caption[]{
\label{matching:fig:unfolding}
\textbf{A \phistructure{} and its intrinsic meaning in a disconnected state.}
\begin{panels}

    \item Illustration of the \phistructure{} specified by the B1 system in its current state, when disconnected from the environment. Subsets of the system (\emph{mechanisms}) that specify an irreducible cause--effect pair (a \emph{distinction}) are plotted as the lowest cluster of points (brown labels), superimposed on the substrate. The causes of each distinction are shown as points in the top-left cluster with purview labels in red; effects are plotted on the top right with purview labels in green (only some labels are shown, for legibility). In each cluster, points are arranged according to a 2D embedding of $n$-dimensional binary vectors that encode the inclusion of units in the mechanism/purview represented by each point (where $n = 13$ is the size of the system). The embedding was computed using UMAP \autocite{mcinnes2020umap} and the same embedding is used for each cluster and for all \phistructure{} visualizations throughout. Red and green lines link the mechanisms to the cause and effect pair they specify. Distinctions are bound together by irreducible \emph{cause--effect relations}, which comprise sets of causes and/or effects that congruently specify common purview units (\emph{relation faces}). Relation faces consisting of two cause and/or effect purviews (\nth{2}-degree faces) are plotted as lines between the points corresponding to those purviews. Higher-degree faces are not shown; this visualization shows only a low-dimensional aspect of the full structure. Mechanisms, purviews, and \nth{2}-degree relation faces are colored according to their $\phi$ value.

    \item Heatmap showing the sum of $\phi$ values for the distinction \phifold{} stemming from each distinction in the \phistructure. \emph{Insets:} Example distinction \phifolds{} superimposed on the background of the full structure in gray. The mechanism (brown point) is shown connected to its cause purview (red point) and effect purview (green point). Relations in the \phifold{} (those involving either the distinction's cause or its effect) are colored as in (A).

\end{panels}
}
\end{adjustwidth}
\end{figure}

To emphasize this point, we first unfold the \phistructure{} of the B1 system in the `dreaming' condition---that is, while it is disconnected from the environment. Fig.\ \ref{matching:fig:unfolding}A shows the results of this analysis for an example state (all units `OFF', inset). The \phistructure{} is composed of 115 distinctions with 54,432 relations among them, with structure integrated information $\Phi = 404.44$.

Fully unfolding the \phistructure{} is computationally infeasible for even this small model system \autocite{mayner2018pyphi}, so for these analyses we only computed a representative subset of the distinctions and relations specified by the system (see \nameref{matching:appendix}). Furthermore, the visualizations of \phistructures{} only plot the mechanisms \& distinctions (as points) and the \nth{2}-degree relation faces that bind two distinctions together (as lines), since these parts of the structure are amenable to two-dimensional representation. However, the \phistructure{} is high-dimensional and also comprises many higher-degree relation faces. Finally, for the purposes of this demonstration, we assume that the system is maximally irreducible; in general, however, the first four postulates of IIT should be applied to all subsets of the substrate and across grains to identify the system that maximizes $\phi_s$.

The $\Phi_d$ values of each distinction \phifolds{} are plotted in \cref{matching:fig:unfolding}B, showing how the structure integrated information $\Phi$ is distributed across the system's irreducible mechanisms. The top inset shows the \phifold{} stemming from mechanism $i$; the bottom inset shows that of mechanism $bcd$ (uppercase and lowercase unit labels denote `ON' and `OFF', respectively). The lattice units in level 1 are designed to exhibit the basic features of a substrate of spatial extendedness as described in \textcite{haun2019why}. Briefly, this is a substrate that specifies a \phifold{} that captures the phenomenal structure of experienced space. As argued in \textcite{haun2019why}, phenomenal space is composed of distinctions (`spots') that are related to themselves (reflexivity) and overlap with other distinctions in a way that satisfies inclusion, connection, and fusion. Similarly, the configuration detectors and segment detector together specify distinctions and relations that capture the notion of an object, binding together an invariant concept and a specific configuration of features (to be discussed in \autocite{grassohow}).

\subsection{The connectedness value and triggering coefficient measure to what extent the state of a complex is triggered by a stimulus}

\begin{figure}[H]
\begin{adjustwidth}{-2.25in}{0in} 
\centering
\includegraphics[width=\maxwidth]{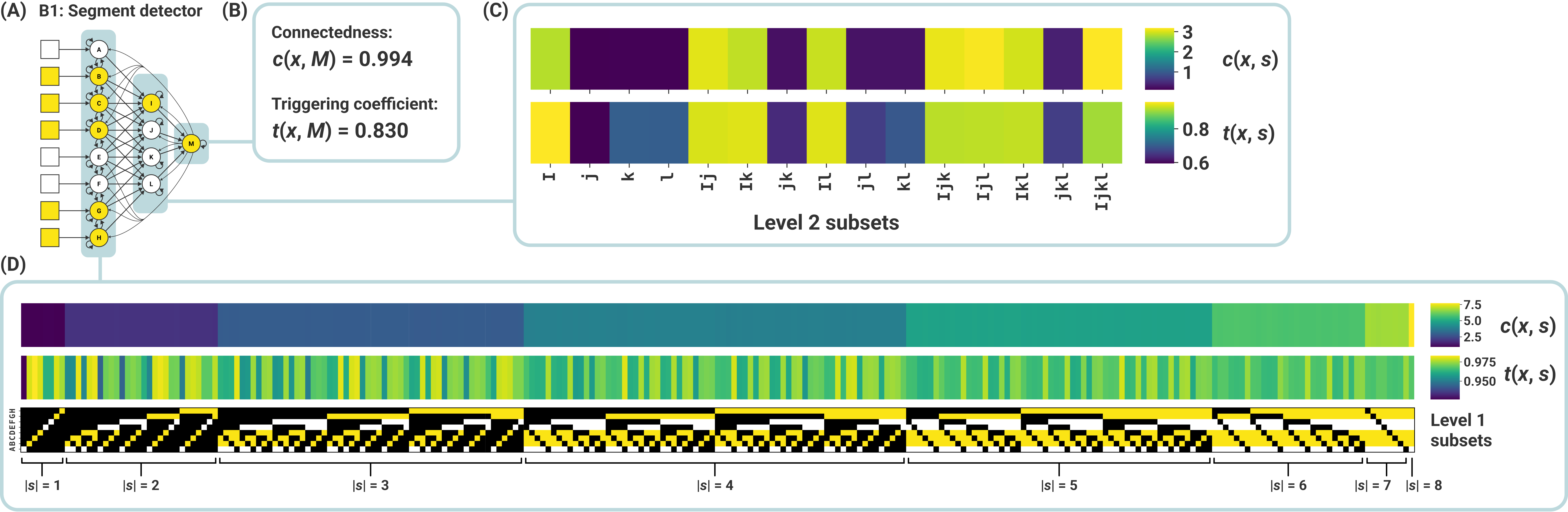}
\caption[]{
    \label{matching:fig:triggering}
    \textbf{Connectedness and triggering coefficients.}
    \textbf{(A)} When the system is connected to the sensory interface, the stimulus $\sensoryinterface = x = 01110011$ percolates through the system, triggering the state $ABCDEFGH = 01110011$, $IJKL = 1000$, $M = 1$ in this example.
    \textbf{(B--D)} Values of $\connectedness(x, s)$ and $\triggering(x, s)$ for this stimulus--response pair are shown for the subsets of levels 3 (B),  level 2 (C), and level 1 (D). Black indicates the unit corresponding to that row is not included; white indicates the unit is included but not active; yellow indicates an included active unit. Due to space constraints, we do not show the values for the full power set, omitting the 7,921 subsets that span levels.
}
\end{adjustwidth}
\end{figure}

When the B1 system is connected to the environment in the `awake' condition, the stimulus percolates through the system and triggers a response. Accordingly, the subsets of the system have positive connectedness values and triggering coefficients, shown in \cref{matching:fig:triggering}.\footnote{Because we have the full TPM $\tpm{U}$ of the substrate in our simulation, the connectedness values do not need to be estimated from repeated trials, as would be necessary in an experiment. Instead, the relevant probabilities can be calculated directly by matrix operations on $\tpm{U}$. First, we compute $\tpm{S}(\sensoryinterface_{t-\tau}) = \Pr(\system_t \given \sensoryinterface_{t - \tau})$ as follows. For each stimulus $\stimulus$, $\tpm{U}$ is conditioned on $\sensoryinterface = \stimulus$ and the resulting conditional TPM of $\system$ is exponentiated by $\tau$ to yield $\Pr(\system_t \given \sensoryinterface_{t - \tau} = x, \system_{t - \tau})$ for each initial system state $\system_{t-\tau} = \responsestate_{t-\tau}$. We then marginalize over $\system_{t-\tau}$ to obtain $\Pr(\system_t \given \sensoryinterface_{t-\tau})$ as desired. Marginalizing over stimuli $\sensoryinterface_{t-\tau}$ then yields $\Pr(\system_t)$. Finally, the probabilities $\conditional$ and $\marginal$ for each subset $M \subseteq \system$ can then be obtained by marginalizing over the appropriate columns and used to compute the connectedness value and triggering coefficient.} Because the stimulus $x = 01110011$ contains the pattern 01110, it causes the configuration detector $I$ and the segment unit $M$ to activate (\cref{matching:fig:triggering}A). $M$ has connectedness $\connectedness(x, M) = 0.994$ and triggering coefficient $\triggering(x, M) = 0.830$, reflecting the fact that its state was caused by the stimulus (\cref{matching:fig:triggering}B). Connectedness values and triggering coefficients for subsets of the level 2 configuration detector units are higher for those subsets that include the activated detector $I$ and lower for those that include only non-active detectors, reflecting the fact that the detectors are strongly and selectively activated by segment states (\cref{matching:fig:triggering}C). For the level 1 subsets, two patterns can be seen. First, connectedness values of level 1 subsets are generally proportional to the size of the subset (\cref{matching:fig:triggering}D). This is because the lattice units in level 1 function largely as feed-forward relays, so the marginal probability of a subset $Z$ being in a particular state is on the order of $1/\card{\Omega_Z} = 1/2^{\card{Z}}$ (since we impose the uniform distribution over stimuli). Second, there is small-scale heterogeneity in connectedness superimposed on the large-scale pattern. This reflects the differential strength of the lateral connections due to short-term plasticity, which is maximal when both units are active. When the connectedness values are normalized to yield the triggering coefficient, the influence of subset size is discounted and the heterogeneity due to the lateral connections dominates. For example, $\triggering(x, C)$ is higher than that of its neighboring units $\triggering(x, B)$ and $\triggering(x, D)$ because $C$ forms the middle of the segment and its co-active neighbors both provide strong lateral input, while $B$ and $D$ are at the border and their respective lateral inputs from $a$ and $e$ are less potentiated. Similar effects can be seen with larger subsets, \eg{} comparing $\triggering(x, BCD)$ to $\triggering(x, aBC)$ and $\triggering(x, CDe)$.

\subsection{A perceptual structure (with the associated value of perceptual richness) is the portion of a \phistructure{} triggered by a stimulus}

\begin{figure}[H]
\begin{adjustwidth}{-2.25in}{0in} 
\centering
\includegraphics[width=\maxwidth]{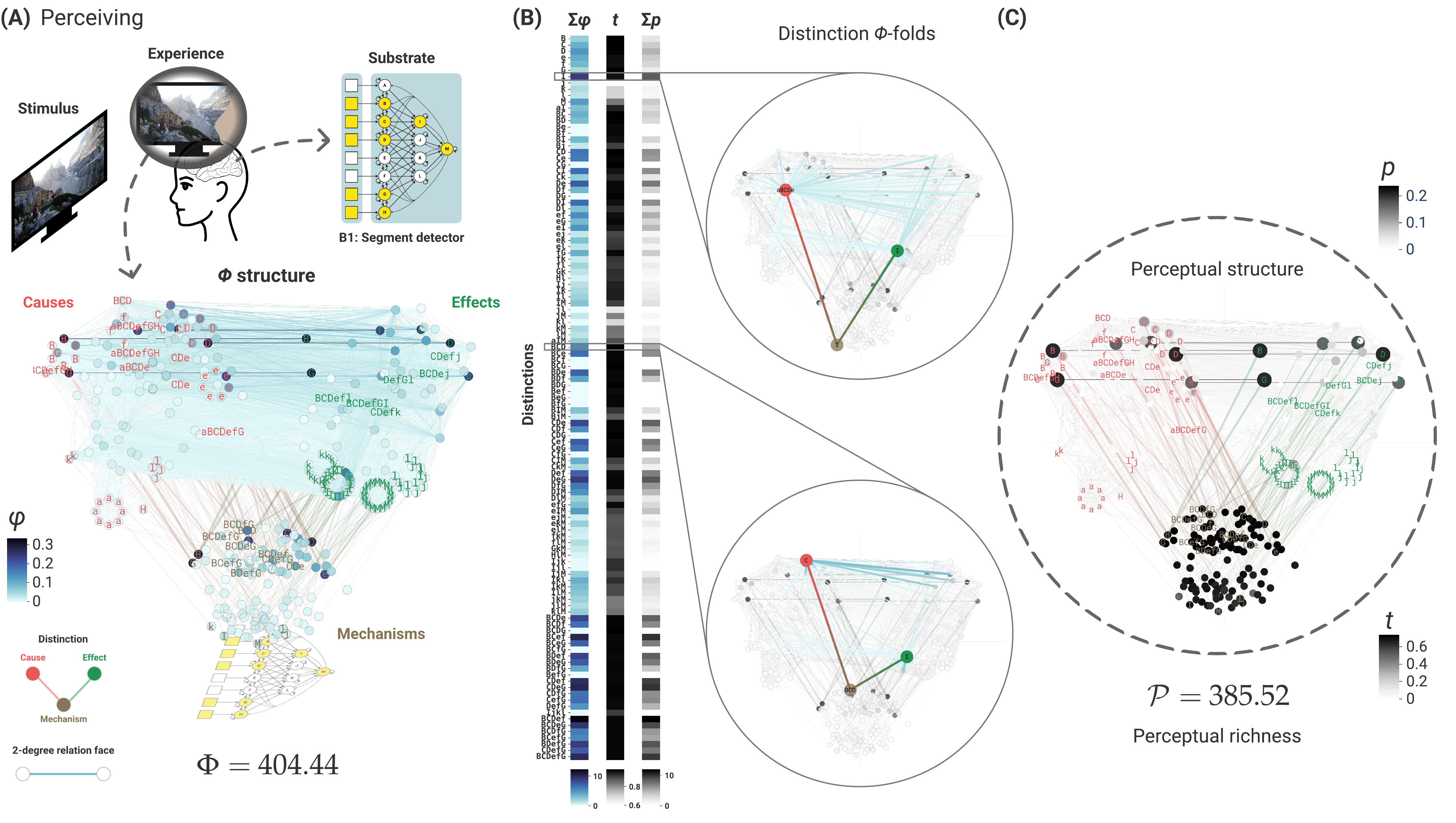}
\caption[]{
    \label{matching:fig:perception}
    \textbf{Perception: perceptual structures and perceptual richness.}
    To the extent that intrinsic meanings are triggered by extrinsic stimuli, they can be considered percepts.
    \begin{panels}

        \item \phistructure{} visualized as in \fig \ref{matching:fig:unfolding}A, with the system now connected to the environment via the sensory interface (`awake' condition). \emph{Inset}: Stimulus and response as in \cref{matching:fig:triggering}; level 1 units relay the stimulus to the levels above, which detect the presence of a segment ($aBCDe = 01110$). Note that in this state, the system specifies a \phistructure{} with higher structure integrated information $\Phi$ than in the all `OFF' state shown in \cref{matching:fig:unfolding}.

        \item Left heatmap and insets as in \fig \ref{matching:fig:unfolding}B. The middle and right heatmaps respectively show the triggering coefficient and perception values associated with each distinction \phifold{}.

        \item The perceptual structure triggered by the stimulus. Points corresponding to irreducible mechanisms $m$ within the system (brown labels, bottom cluster) are colored according to their triggering coefficient $\triggering(\stimulus, m)$. The causes (red labels, top-left cluster), effects (green labels, top-right cluster), and relations among them (lines within and between clusters) specified by the irreducible mechanisms are colored according to their perception value $\perception(\stimulus, d(m)) = \triggering(\stimulus, d(m))\;\phi_d(m)$. Grayscale is used to emphasize that the perceptual structure does not exist in its own right as a \phistructure{}, but rather is the fraction of the \phistructure{} that is triggered by a stimulus.

    \end{panels}
}
\end{adjustwidth}
\end{figure}

As in the dreaming condition, the system specifies a \phistructure{} intrinsically (\cref{matching:fig:perception}A).\footnote{By definition, the sensory interface constitutes the background conditions in IIT's analysis (see \S~\nameref{matching:sec:theory:overview}). Therefore, when unfolding the \phistructure{}, we condition the system's TPM on the state of the sensory interface.} The leftmost plot in \cref{matching:fig:perception}B shows how the structure integrated information $\Phi$ is partitioned into the $\Phi_d$ values of the distinction \phifolds{} specified by the system's irreducible mechanisms.

Note that $\Phi$ tends to be concentrated in the distinction \phifolds{} of mechanisms that include the lattice units $BCD$, the corresponding detector unit $I$, or the segment unit $M$. For example, the \phifold{} of the \nth{1}-order mechanism $I$ contributes a relatively large amount to the total $\Phi$. This is partly due to the convergence of many irreducible effects of other mechanisms onto $I$ (note the density of $I$ labels in the effect cluster of the \phistructure{} in panel A). Because $I$ specifies its own state as its maximally irreducible effect, this convergence in turn leads to a high density of irreducible relations among the mechanisms whose effect purviews include $\{I\}$ (these relations contribute $\sum \phi_r = 8.79$ to the $\Phi_d(I) = 8.88$, whereas for the distinction \phifold{} stemming from $j$, for example, the corresponding relations contribute $\sum \phi_r = 0.981$ out of $\Phi_d(j) = 1.01$).

The triggering coefficients $\triggering(x, m)$ and perception values $\Perception_d(x, m) = \triggering(x, m)\;\Phi_d(\ces(d(m))$ of the distinction \phifolds{} are shown in the middle and rightmost plot in \cref{matching:fig:perception}B, respectively. As seen in \cref{matching:fig:triggering}, the triggering coefficients are large for mechanisms that include the active units involved in detecting the segment: $B, C, D, I,$ and $M$. The distribution of $\Perception_d(m)$ values indicates that the intrinsic meanings defined by certain \phifolds{}---such as those of $I$, $CD$, $BDe$, $CDe$, and of higher-order mechanisms that include combinations of $B, C,$ and $D$---can be largely attributed to the stimulus's triggering action and are thus highly perceived.
\cref{matching:fig:perception}C illustrates the resulting \textit{perceptual structure}---the portion of a \phistructure{} triggered by the stimulus. The perceptual structure has an associated value of perceptual richness $\Perception$. The depiction in the figure is necessarily partial---higher-degree relations are omitted as they are difficult to visualize due to their density, even though, as mentioned earlier, they contribute significantly to perceptual richness $\Perception$. Nonetheless, one can see, for example, that mechanisms $B$, $C$, $D$, and $BCD$ specify distinctions with relatively large values of $\perception$, and each distinction's cause and effect are bound by a relation with a perception value as well (color of lines connecting purviews).

Note that in the model, all triggering coefficients are positive, so the perceptual structure includes all components of the \phistructure{}. In general, however, a component of the \phistructure{} may have a triggering coefficient of zero (indicating it was not caused by the stimulus), and in that case it would not appear in the perceptual structure at all. Conversely, if a subset does not specify an irreducible cause and effect within the system ($\phi = 0$) or does so by specifying a state that is incongruent with the state specified by the complex as a whole, it would likewise not appear in the perceptual structure, despite having a positive triggering coefficient. This implies that even units within the neural substrate of consciousness may be activated by a stimulus and yet not contribute to experience, as may be the case during bistable perception or during certain stages of sleep.

\subsection{The same stimulus can trigger different perceptual structures in different systems: every perception is an interpretation}

According to IIT, the meaning of an experience is intrinsic: it is wholly accounted for by the \phistructure{}, which is determined by the complex's connectivity and the cause-effect power of its elements in their particular state, regardless of how it was triggered. Thus, the meaning of the perceptual structure---the portion of the \phistructure{} that was triggered by a stimulus---is likewise wholly intrinsic to the complex. In this sense, \emph{every perception is an interpretation}.

\begin{figure}[H]
\begin{adjustwidth}{-2.25in}{0in} 
\centering
\includegraphics[width=\maxwidth]{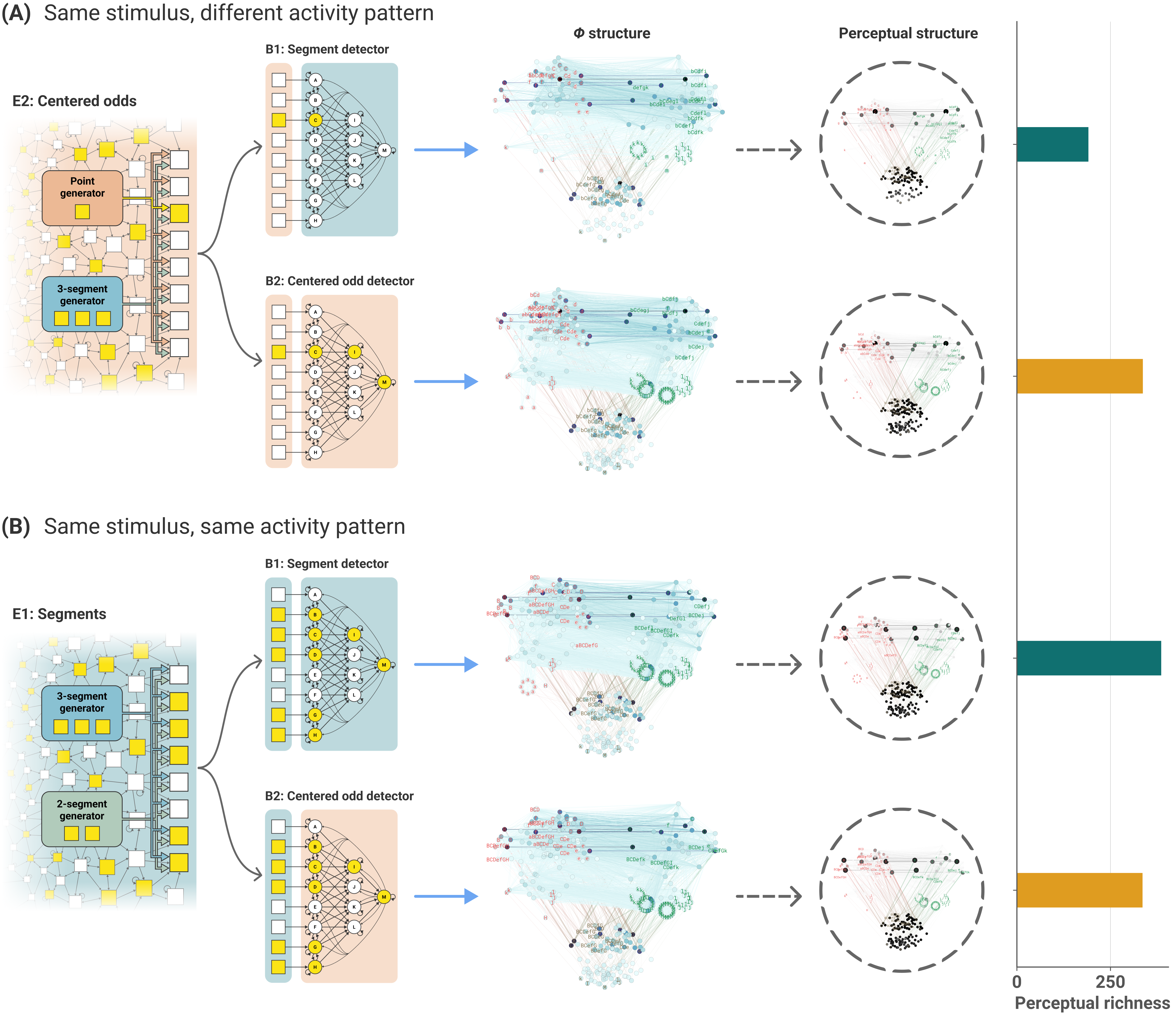}
\caption[]{
    \label{matching:fig:interpretation}
    \textbf{Perception as interpretation.}
    The same stimulus can trigger different perceptual structures in different systems. Importantly, this can happen even when the activity pattern in both systems is the same.
    \begin{panels}
        \item B1 and B2 are presented with the stimulus 00100000, which contains the centered odd pattern `00100', one of the configurations B2 is designed to detect. Accordingly, the perceptual structure triggered in B1 has a lower perceptual richness than that triggered in B2.
        \item In this case, the stimulus presented is preferred by both B1 and B2---it contains '01110', both a segment and a centered odd pattern---and the top level invariant detector is activated in each system. In fact, the activity pattern triggered in the two systems is identical. However, even though the stimulus and triggered activity are the same, the perceptual structures---and therefore, by IIT, the intrinsic meanings of the stimulus---are different, emphasizing that activity patterns do not have meaning in themselves. Here, the only difference between the two situations is the \emph{counterfactual} behavior of the L2 configuration units---that is, a difference in the \emph{causal powers} of those units. In this sense, each system interprets the stimulus according to its causal powers.
    \end{panels}
}
\end{adjustwidth}
\end{figure}

To illustrate, consider systems B1 and B2 exposed to two different environments (\cref{matching:fig:interpretation}). The `segment' environment, E1, consists of a 3-segment generator and a 2-segment generator against a low-level noise background. Each generator causes a number of contiguous units on the sensory interface to activate at a uniformly random location (3 contiguous units with probability $p = 0.6$ and 2 contiguous units with probability $p = 0.9$, respectively). The `centered odd' environment, E2, contains a 3-segment generator and a point generator (which activates a single unit), with respective probabilities $p = 0.6$ and $p = 0.9$. In each environment, the noise background causes each sensory interface unit to activate with probability $p = 0.05$.

First, a 00100 pattern is presented to both B1 and B2 (\cref{matching:fig:interpretation}A). By construction, this pattern does not ```resonate'' well with the B1 system, since it is not a segment and does not activate the higher level detectors and invariant units. In the B2 system, the stimulus triggers a perceptual structure with higher perceptual richness, as the cause-effect power of B2's units was designed to be sensitive to this pattern; to the B2 system, this stimulus means that a centered odd pattern is present at a particular location, in the sense that the invariant unit $M$ and the configuration unit $I$ are coactivated and specify a corresponding set of \phifolds{}.

Importantly, the same stimulus can trigger different perceptual structures in different systems even if the systems' activity patterns are exactly the same. In \cref{matching:fig:interpretation}B, a 3-segment stimulus is presented to B1 and B2. Since this stimulus is both a segment and a centered odd, the same response state is triggered in B1 and B2, with the $I$ and $M$ units active in both systems. However, the triggered \phistructure{} and perceptual structures differ, particularly the \phifolds{} specified by $I$ and $M$. The B1 system interprets the stimulus as a segment (and not a centered odd) while B2 interprets it as a centered odd (and not a segment), emphasizing again that in this account it is the causal power of the system's units that matters for perception, rather than activity patterns as such.

\subsection{Perception may or may not represent causal features of an environment, but is likely to do so under adaptive constraints}

\begin{figure}[H]
\begin{adjustwidth}{-2.25in}{0in} 
\centering
\includegraphics[width=\maxwidth]{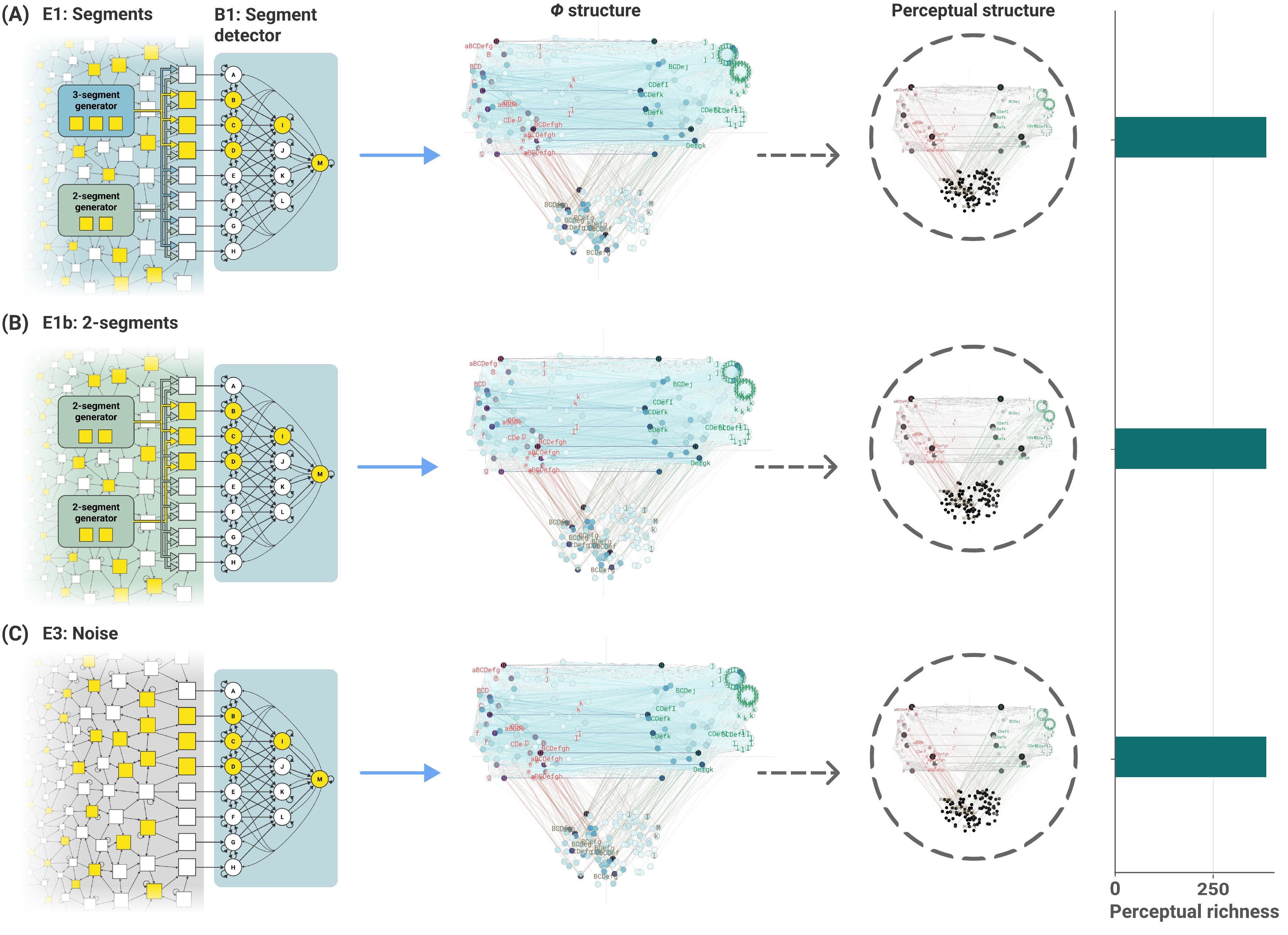}
\caption[]{
    \label{matching:fig:representation}
    \textbf{Perception and representation.}
    \begin{panels}
        \item The B1 system is exposed to the segment environment E1 and detects the 3-segment pattern in the stimulus.
        \item In the `2-segment' environment E1b, the 3-segment generator is replaced with a 2-segment generator. With two 2-segment generators, some combinations of 2-segment locations result in the 2-segments overlapping on one sensory interface unit, forming an `apparent 3-segment'. When the segment detector system is exposed to the 2-segment environment and perceives an apparent 3-segment stimulus, the 3-segment has the same meaning for the system as when it is generated by the segment environment (bottom). And yet in the first case, by construction, there is no causal process in the 2-segment environment to which the 3-segment percept can refer or 'represent': the apparent 3-segments are 'spurious coincidences', conditional on the presence of 2-segments.
        \item In the noise environment E3, the 3-segment pattern can occur purely by chance. As in the E1b case, the system responds in the same manner, and the 3-segment pattern has the same meaning for the system. Here there are no causal processes whatsoever in the environment, so again, there is nothing that the 3-segment percept can represent or refer to.
    \end{panels}
}
\end{adjustwidth}
\end{figure}

To illustrate the implications of our formalism for notions of representation and reference, we introduce two additional environments, E1b and E3. In the E1b environment, the 3-segment generator has been replaced with a 2-segment generator. Occasionally, the 2-segments overlap such that they form a 3-segment (\cref{matching:fig:representation}A, left).  Conditional on the occurrence of the 2-segments, these `apparent 3-segments' occur at chance levels.  The E3 environment consists of pure noise, such that all possible stimuli occur with uniform probability.

When the B1 system is connected to E1b and an apparent 3-segment occurs (\cref{matching:fig:representation}B), the system naturally responds in the same manner as when it is presented with a `true' 3-segment generated by E1, the segment environment (\cref{matching:fig:representation}A). Likewise, B1 responds in the same manner when a 3-segment occurs purely by chance in the noise environment E3 (\cref{matching:fig:representation}C). In the example shown, the response states $y$ are the same in each case, so the system unfolds into the same \phistructure{} and perceptual structure, regardless of whether the corresponding experience is a 3-segment that was generated by some causal feature in E1, merely a confluence of 2-segments that form an apparent 3-segment in E1b, or a purely random pattern in E3.

This demonstrates a simple but important point. In the E1 case, there is a causal process in the environment---a 3-segment generator---that causes the 3-segment to appear above chance levels, and which the resulting perceptual structure can be said to ``represent''; that is, the content of the experience triggered by the stimulus---the percept---has a true, inter-subjective referent in the environment. By contrast, in the E1b case, there is no such causal feature, since by construction, the only causal features of the environment are 2-segment generators. The apparent 3-segment is, in a sense, a `trick of the light': once the actual causal processes in the environment are taken into account, the 3-segments are revealed to be spurious coincidences. In the E3 case, there are no causal features whatsoever; the environment is completely noisy. Thus in E1b and E3, there is no referent in the environment to which the percept can refer, no causal feature that can be ``represented''. And yet, by IIT, the experiences triggered in all three cases are identical.

The point here is not to highlight the difficulty of the problems of causal inference or generative model selection; rather, it is to emphasize how, in IIT's account, perception and meaning, being intrinsic, do not necessarily involve reference and representation of causal features of the environment.

In the E1b case, while the intrinsic meaning ``3-segment'' does not refer to a causal process in the environment, it does nevertheless capture a statistically relevant feature: 3-segments do indeed occur more frequently than would be expected in the pure noise environment. More generally, however, intrinsic meanings, even when triggered by external stimuli, may be idiosyncratic, arbitrary categorizations.

That said, as will be discussed later, the waking experiences of evolved organisms that have been exposed to sequences of stimuli within a given environment can be expected to be related to causal features of the environment, and that relationship can in many cases be usefully described as ``representation,'' although the mapping is generally far from straightforward.

\subsection{Perceptual differentiation measures the richness and diversity of perceptual structures triggered by a sequence of stimuli sampled from an environment}

\begin{figure}[H]
\begin{adjustwidth}{-2.25in}{0in} 
\centering
\includegraphics[width=\maxwidth]{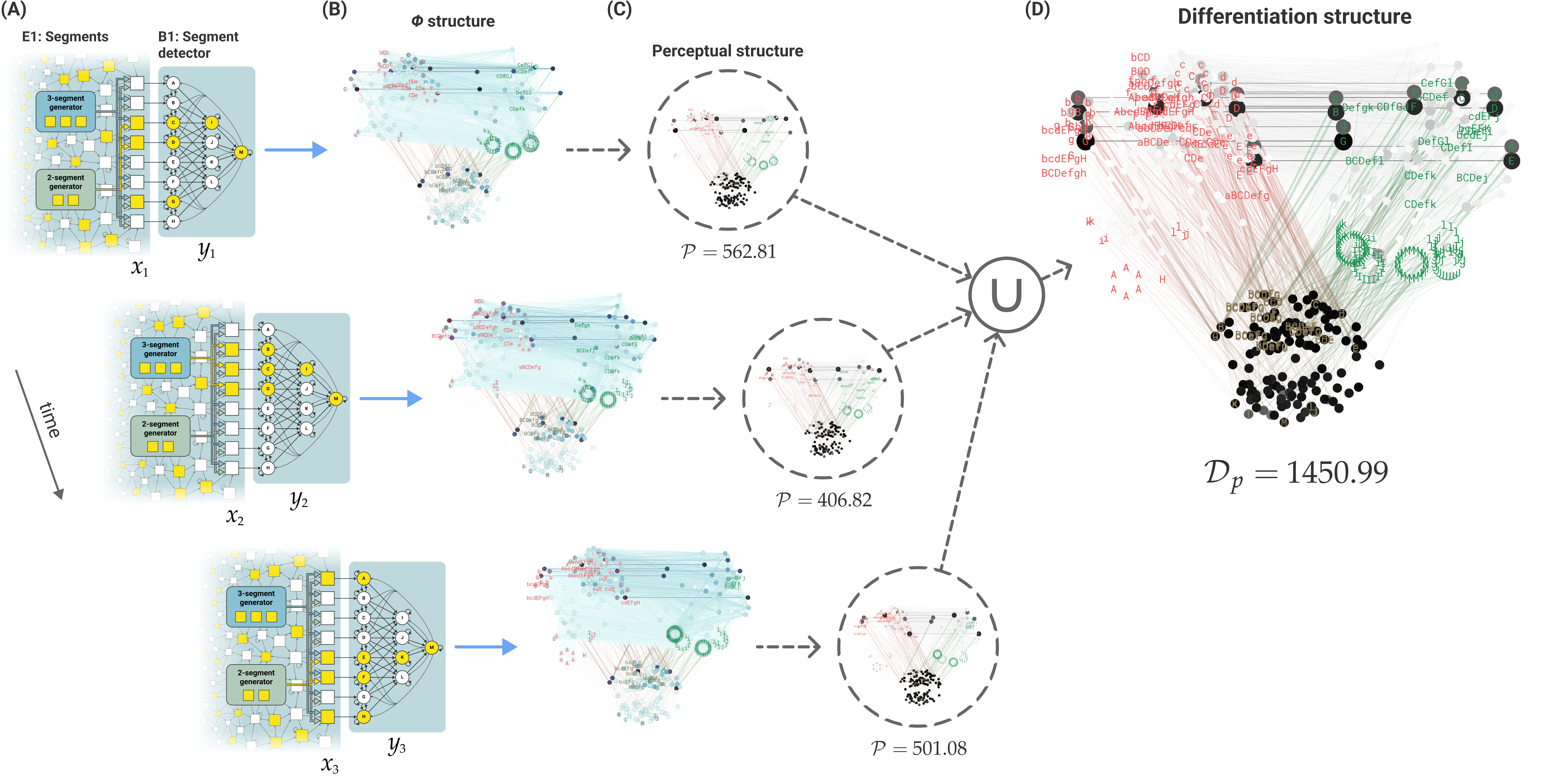}
\caption[]{
    \label{matching:fig:differentiation}
    \textbf{Perceptual differentiation.}
    \begin{panels}
        \item A sequence of stimuli is presented to the system.
        \item Each stimulus triggers a \phistructure{} composed of distinctions and relations.
        \item When the $\varphi$ values of these components are weighted by their associated triggering coefficients, we obtain a perceptual structure, which expresses the portion of the \phistructure{} that was triggered by the stimulus.
        \item Taking the union of these structures---\ie, counting only once the distinctions and relations that are triggered by multiple stimuli in the sequence---yields a \emph{perceptual differentiation structure}. The sum of the $\varphi$ values of its components is the \emph{perceptual differentiation} triggered in this system by the stimulus sequence.
    \end{panels}
}
\end{adjustwidth}
\end{figure}

Now we consider sequences of stimuli impinging on the system from the environment, rather than a single stimulus in isolation. A sequence of stimuli will trigger a corresponding sequence of perceptual structures, which will generally vary in their perceptual richness (\cref{matching:fig:differentiation}A--C). Furthermore, certain distinctions or relations within those structures may be triggered repeatedly.

We measure the diversity of perceptual structures triggered by a set of stimuli with \emph{perceptual differentiation}: the sum of the perceptual richness of each triggered perceptual structure, where the $\varphi$ contributed by distinctions and relations that appear as components of multiple structures across the set are only counted once. The union of the distinctions and relations triggered by the sequence is called the \emph{perceptual differentiation structure} (\cref{matching:fig:differentiation}D).

Perceptual differentiation is high if the stimulus sequence triggers structures that both (1) have high perceptual richness and (2) are different from one another.\footnote{Note that repeated presentations of the same stimulus may nonetheless trigger a differentiated set of perceptual structures if the system's state depends on previous stimuli, \ie, if it has some form of memory.} In this sense, it measures meaningful the stimulus sequence is to the system.

\subsection{Matching: perceptual differentiation triggered by representative samples from an environment measures the intrinsic meaningfulness of that environment to a complex}

\begin{figure}[H]
    \begin{adjustwidth}{-2.25in}{0in} 
    \centering
    \includegraphics[width=\maxwidth]{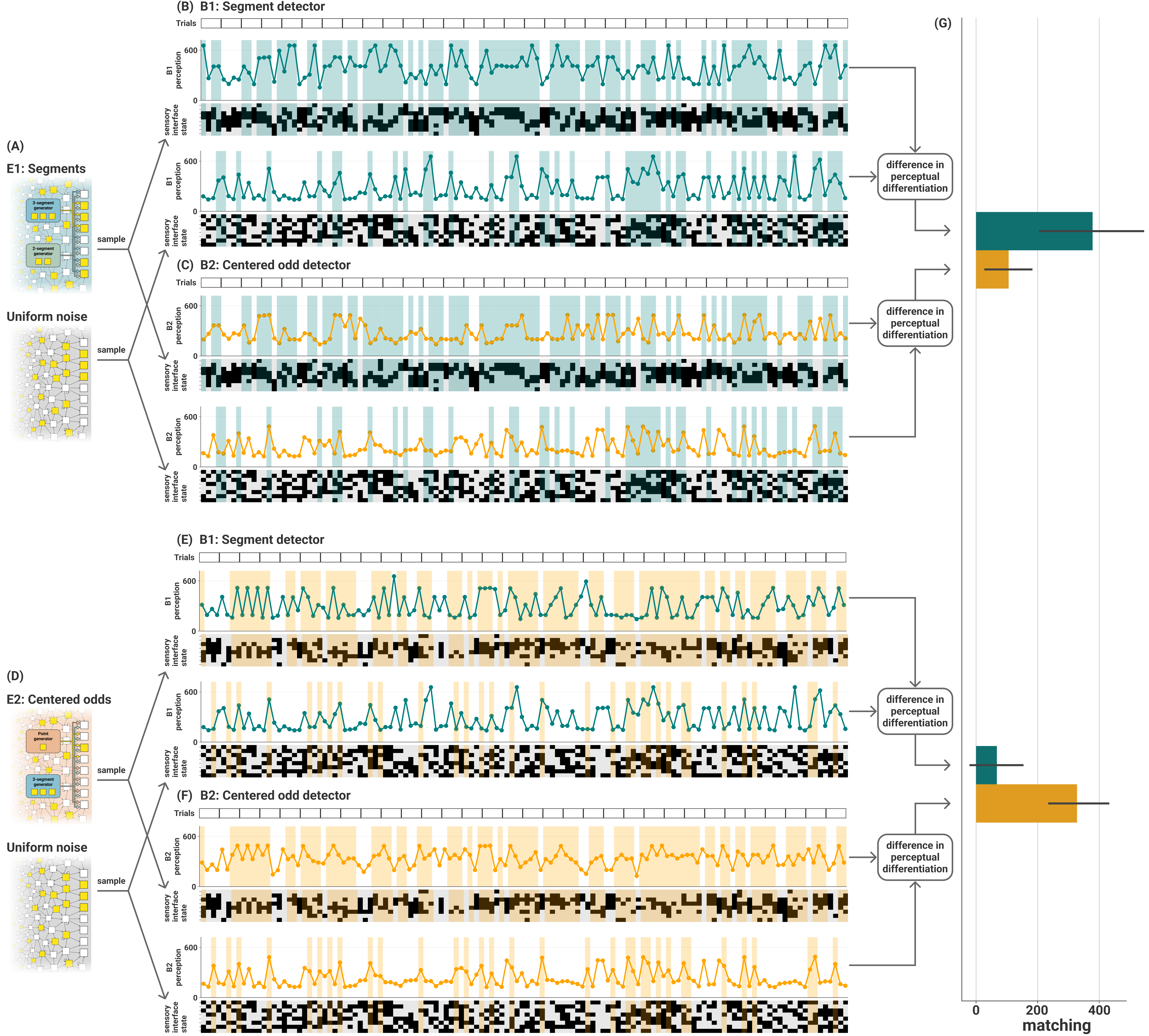}
    \caption[]{
        \label{matching:fig:matching}
        \textbf{Matching in different environments.}
        \textbf{(A)} B1 and B2 are exposed to the segment environment E1 (top) and to uniform noise (bottom). $n = 32$ trials of sequences of $k = 4$ stimuli were sampled from each.
        \textbf{(B)} Perceptual richness $\Perception$ triggered in B1 by each environmental stimulus (top) and random stimulus (bottom). Samples that included any segment pattern (2-segments or 3-segments) are highlighted in teal.
        \textbf{(C)} Same as (B), but for B2. Note that the segment stimuli tend to trigger higher perceptual richness in B1 than in B2, since B1 is designed to detect any segment pattern, while B2 can only detect 3-segments.
        \textbf{(D--F)} Same as (A-C), but for the centered odd environment E2. Stimuli containing a centered odd pattern are highlighted in orange. Here, the stimulus sequences tend to trigger greater perceptual richness in B2 than in B1, as B2 can detect both 3-segments and the single-dot pattern $00100$, while B1 can only detect the 3-segments.
        \textbf{(G)} Matching is estimated as the maximum mean difference between perceptual differentiation triggered by environmental stimulus sequences and random stimulus sequences, where the maximum is taken over contiguous subsequences of length $l \leq k$ within each trial, and the mean is taken across trials (\cref{matching:sec:theory:matching}).
        As expected, B1 matches better to E1 and B2 matches better to E2, reflecting that the each system has, by construction, ``internalized'' aspects of the stimulus statistics that in turn reflect causal processes in their ``matching'' environments. Two-way ANOVA system $\times$ environment interaction: $F(1, 124) = 20.23$, $p < 0.001$; no significant main effects. Post-hoc Tukey tests: B1:E1 \textgreater B2:E1, $p=0.008$; B1:E1 \textgreater B1:E2, $p=0.002$; B2:E2 \textgreater B1:E2, $p=0.012$; B2:E2 \textgreater B2:E1, $p=0.044$. Error bars represent 95\% confidence intervals.
    }
    \end{adjustwidth}
\end{figure}

Matching is formally defined as the maximum expected difference between the perceptual differentiation triggered in a complex by sequences sampled from an environment and that triggered by uniformly random stimulus sequences, for a given stimulus sequence length $k$ (where the maximum is taken over contiguous subsequences of length $l \leq k$). \cref{matching:fig:matching} shows the values of perceptual richness $\Perception$ triggered in systems B1 and B2 across $n = 32$ trials of $k = 4$ sequential stimuli sampled from environments E1 (generating segment stimuli), E2 (generating centered odd stimuli), and from the uniform distribution (the noise environment). Stimuli that are much more frequent than would be expected by chance are highlighted (\ie, those containing a segment or a centered odd). Matching is then estimated as the maximum mean difference between the perceptual differentiation triggered by the environmental sequences and that triggered by the random sequences, where the mean is taken over $n = 32$ trials and the maximum is taken over contiguous subsequences of the $k = 4$ stimuli in each trial (\cref{matching:fig:matching}G).

As expected, the matching value is higher for B1 with E1 and for B2 with E2, reflecting the fact that their wiring has ``internalized'' aspects of the stimulus statistics that in turn reflect causal features of E1 and E2, respectively. This is because (1) in each system's ``matching'' environment, preferred stimuli occur more frequently; (2) in each system, the preferred stimuli yield higher perceptual richness, as can be seen in the colored regions in the perceptual richness plots (\cref{matching:fig:matching}B,C,E,F); and (3) different preferred stimuli yield perceptual structures that differ more than those triggered by non-preferred stimuli.

In general, there will be a stimulus sequence length $k$ that maximizes the matching value for a given system in a given environment. However, in realistic systems with large repertoires of sensory interface states and internal states, the optimal $k$ may be impractically long, and achieving it may require impractically large numbers of trials. Since both the length and number of trials in experiments are necessarily limited, this places a premium on choosing representative stimulus sequences based on heuristic criteria that are informed by existing knowledge of the environment and of the system under study.

To the extent that the stimulus sequences are representative of a system's environment, the matching value can serve as a measure of how richly and diversely the system interprets inputs from that environment, reflecting how well it ``resonates'' with it. Like perception, perceptual differentiation and matching do not necessarily capture causal features, and typically not in a straightforward manner.

Nevertheless, for systems that need to adapt to complex environments, we can expect that through evolution, development, and learning, causal features of the environment that are relevant for survival will lead to adaptive changes in their intrinsic causal organization. This would typically result in high values of matching, with the intrinsic meanings that are available to the system---the distinctions and relations supported by its units---allocated preferentially to the interpretation of important causal features of the environment.

\section{Discussion}
\label{matching:sec:discussion}

IIT argues that every experience is accounted for by a \phistructure{} specified by a complex in a state. The \phistructure{} captures in full the way the experience feels, which is the same as its intrinsic meaning \autocite{albantakis2023integratedb, tononi2022only}. This paper extends the IIT framework by presenting a principled approach to quantify the triggering of intrinsic meanings by stimuli sampled from an environment. Below we briefly review the framework presented through minimalistic models in the \nameref{matching:sec:theory} and \nameref{matching:sec:results} sections and examine some of its implications.

\paragraph{Experience and intrinsic meaning.}~

The theoretical framework presented here starts deliberately from the intrinsic meaning of an experience. According to IIT, the essential properties of consciousness---those that are true of every conceivable experience---are as follows:  every experience is intrinsic (for the subject), specific (this one), unitary (a whole, irreducible to its parts), definite (this whole, containing all it contains, neither less nor more), and structured (it feels the way it feels). An experience is fully characterized in quantity and quality, with no additional ingredients, by the \emph{cause-effect structure}, or \emph{\phistructure{}} specified by the mechanisms of a complex in its current state. For example, to account for how the experience of “seeing a segment” feels, a complex must specify a \phistructure{} whose causal distinctions and relations are structured just like the experience: some must account for the feeling of spatial extendedness \autocite{haun2019why}, others for the feeling that the general concept “segment” is bound to a particular configuration \autocite{grassohow}, and others for the feeling that the configuration is located at a particular region of space. What the experience means for the subject is the same as what it feels like---\emph{“the meaning is the feeling”} \autocite{tononiforthcomingbeing}. Therefore, the meaning of an experience, like its feeling, is intrinsic, specific, unitary, definite, and structured. The intrinsic nature of experience also implies that its feeling/meaning is the same whether the experience occurs spontaneously, as in a dream, or it is triggered by stimuli from the environment. In fact, phenomenally it may at times be difficult to tell whether a particular scene is dreamt or triggered by extrinsic stimuli \autocite{nir2010dreaming,wamsley2014delusional}.

The quantity or richness of feeling/meaning is measured by \emph{structure integrated information} ($\Phi$), which is the sum of the integrated information values $\phi$ of all the distinctions and relations that compose the \phistructure{}. The quality of the experience is given by the distinctions and relations that compose the \phistructure{}, with no additional ingredients \autocite{albantakis2023integratedb}. The contribution to feeling/meaning of every subset of the complex is given by its distinction \phifold{}. This is the sub-structure composed of the distinction specified by the subset and the associated relations. Its $\Phi_d$ value is the sum of the distinction's $\phi_d$ value and the $\phi_r$ values of each associated relation, divided by the number of distinctions which that relation binds together.

In Fig. \ref{matching:fig:model}, this is illustrated by considering a minimalistic model of a sensory hierarchy in which a pattern of activity over a “macro” time scale (corresponding, say to tens of milliseconds) has been established through internal interactions at a finer time scale (corresponding, say to a few milliseconds). According to IIT, as long as the neural substrate is maximally irreducible at the macro time scale (\ie, it is a complex), it specifies a \phistructure{} whose distinctions and relations compose its intrinsic feeling/meaning. Thus, L1 units specify “spots” bound by relations (overlaps among distinctions’ causes and effects) of reflexivity, inclusion, connection, and fusion, into a spatial sub-structure \autocite{haun2019why}; L2 units specify “configurations” of L1 units bound to that spatial sub-structure; and L3 units specify “invariants” bound to a specific configuration and its spatial sub-structure \autocite{grassohow}. This holds whether the current activity pattern came about intrinsically, as in a dream, or was triggered by a sensory stimulus.

\paragraph{Connectedness and triggering coefficients.}~

As defined here, connectedness assesses the extent to which the state of subsets of units within a system is \emph{caused} by a stimulus (the state of units of the sensory interface). The measure is based on the actual causation formalism, which employs the principles of IIT to quantify ``what caused what'' \autocite{albantakis2019what}. Subset connectedness is high if the probability of occurrence of the current subset state is increased by the stimulus compared to all possible stimuli. The normalized value\footnote{The normalization is by the logarithm of the reciprocal of the probability of that state, known as self-information.} of subset connectedness is the \emph{triggering coefficient}. It varies between 0, if a stimulus makes no difference to the probability of occurrence of that subset state (say, owing to a disconnection of sensory pathways), and 1, when that subset state always occurs in response to that stimulus.

Phenomenally, when we are awake, our ongoing experiences are typically triggered by external stimuli (though at times they may be ignored during states of absorption or mind-wandering). When we are asleep, especially during rapid eye movement (REM) sleep, the same stimuli fail to reliably trigger any experience, which proceeds on its own track as a dream sequence (though at times a stimulus may trigger some dream content that is “incorporated” in the dream narrative \autocite{tononi2024consciousness}).

In the model, connected to its environment through the sensory interface, stimuli trigger an activity pattern in L1, which rapidly percolates to L2 and L3. This is meant to resemble what happens in early visual hierarchies of the brain during wakefulness \autocite{riesenhuber1999hierarchicala}. The model is designed such that the stimulus conveyed by the sensory interface activates the appropriate ports-in units in L1 at a fast time scale (a “micro” time step). The activation quickly percolates to L2 and L3, mimicking the fast “feed-forward sweep” in the brain \autocite{lamme2000distincta}. The complex “endorses” the state triggered by the stimulus by amplifying the efficacy of neuronal interactions for tens to hundreds of milliseconds (a ``macro'' time step). This intrinsic endorsement is compatible with the classic notion that the intrinsic connectivity ``amplifies'' and sustains the initial effects of sensory inputs through various specializations \autocite{chance1999complex,douglas1995recurrent,li2013intracortical,lien2013tuned,peron2020recurrent,pattadkal2022primate}. However, amplification is usually conceived in terms of increased gain of neural activation/deactivation, whereas the model emphasizes the role of a transient increase in connectivity strength---\ie, short-term plasticity. The endorsement of the activity pattern triggered by the stimulus allows extrinsic connections to effectively drive cortical activity, preventing the occurrence of ``hallucinatory'' patterns, despite the much larger number of intrinsic connections.\footnote{Notably, in sensory regions of the brain the connectivity intrinsic to each area far outweighs the extrinsic connectivity originating from subcortical inputs or other cortical areas, except for directly adjacent areas \autocite{binzegger2004quantitative, binzegger2009topology, vezoli2021cortical}. For example, 85--90\% of connections to primary visual cortex originate within it, but less than 1\% originate in the lateral geniculate nucleus of the thalamus, its primary sensory afferent \autocite{binzegger2004quantitative, markov2011weight}. The intrinsic connectivity is extraordinarily dense locally (within 1--2 millimeters) and especially so within supragranular layers \autocite{binzegger2004quantitative, binzegger2009topology}. The latter are especially well represented in much of central/posterior cortex \autocite{castrillon2023energy}. Moreover, posterior cortex has high cell density---neurons tend to be small and are more likely to have short axons, privileging local connectivity. Prefrontal cell density is lower, \ie, cells are on average larger, which may allow them to support longer axons that can connect distant cortical areas \autocite{goulas2019blueprint} or reach outside the cortex. Crucially, in much of posterior cortex, the intrinsic connections among specialized units are organized as a lattice, which fits the requirement of maximal integration for the substrate of consciousness \autocite{tononi2016integrated, albantakis2023integratedb}, whereas many prefrontal regions tend to be organized in a modular manner \autocite{watakabe2023local}.}
At the same time, it enables a rapid, non-linear selection of activity patterns that resonate with the connectivity, which in turns reinforces the winning patterns.\footnote{Minor refinements of noisy activity patterns are compatible with this scenario. Moreover, the progressive recruitment of intrinsic “loops” of increasing length \autocite{hashmi2013sleepdependent} enables a “compositional search” where low-order (single-neuron), local congruence (between feed-forward and feed-back or lateral inputs) is progressively complemented by higher-order (subsets of multiple neurons), global congruence.}
Finally, the marked increase in the efficacy of intrinsic interactions ensures that the maximum of cause-effect power is achieved intrinsically, by the complex over itself, at the macro time scale that corresponds to that of experience \autocite{albantakis2023integratedb}.

Note that the strength of endorsement in the model is maximal when both pre- and post-synaptic units are active. This mimics the action of NMDA receptors, especially concentrated in supragranular layers \autocite{zilles2017multiple}, which can multiply the strength of activated synapses \autocite{mayer1984voltagedependent,nowak1984magnesium} (\eg \autocite{sabatini2002life,nevian2004single}), of metabotropic glutamate receptors, as well as the effects of dendro-somatic coupling \autocite{larkum1999new,larkum2004topdown}. In this way, active units contribute more than inactive ones to the content of an experience (by specifying distinction \phifolds{} of higher $\Phi_d$). This fits with the notion that in the brain, due to energy constraints, strong activation should be reserved for the selective signaling of highly meaningful stimuli (a face, an object, and so on) that need to percolate deeply within the system \autocite{balduzzi2013what}.

\paragraph{Perception as interpretation.}~

A \textit{perceptual structure} is defined here as the portion of the \phistructure{} triggered by a stimulus. \textit{Perceptual richness} can be quantified as the sum of the $\phi$ values of all the distinctions and relations triggered by a stimulus, weighted by their triggering coefficient (or equivalently, the sum of the weighted $\Phi_d$ values of each distinction \phifold{}). Because the meaning of an experience is wholly intrinsic---it is provided by the \phistructure{} regardless of whether or how it was triggered---a perceptual structure can be thought of as an \emph{interpretation} of a stimulus by the complex in its current state. In this sense, \emph{perception is interpretation}.

The model illustrates how perceptual structures and perceptual richness can vary with different stimuli. By measuring how much intrinsic meaning is triggered by a stimulus, perceptual richness quantifies its overall meaningfulness to the complex. Thus, system B1 “resonates” well with stimulus 0111011, which it interprets as a ``segment'' with a particular configuration at a particular location in space (\cref{matching:fig:interpretation}B, top). Other stimuli, such as 0010000, resonate less, triggering a smaller and weaker set of distinctions and relations, because they are not interpreted as belonging to a category available to the complex (\cref{matching:fig:interpretation}A, top). By contrast, 0010000 is highly meaningful to system B2, triggering a larger and stronger set ((\cref{matching:fig:interpretation}A, bottom). The model also illustrates how the same stimulus can trigger different percepts in different systems: in \cref{matching:fig:interpretation}B, system B1 interprets 0111011 as a `segment,' while system B2 interprets it as a `centered odd.' In this case, even the activity pattern is the same, highlighting again that the \phistructure{} specified by an activity pattern over a complex can account for feeling/meaning, while an activity pattern as such has no meaning in itself.

Phenomenally, we can easily recognize this situation by considering how we experience a frame of a movie versus a typical ``TV noise'' frame. The former, which percolates deeply within the brain \autocite{mensen2018differentiation, mensen2017eeg, boly2015stimulus, mayner2022measuring}, triggers an experience rich in structure, with many meaningful, high-level interpretations bound to low-level contours and features. The latter, which percolates much less, affecting mostly primary sensory areas, will trigger an experience whose structure will be almost exclusively spatial, with hardly any high-level meaning, except perhaps for the concept of ``noise.'' Note also that, consistent with IIT, we cannot help interpreting a sensory input as spatially organized even if its source is purely random \cite{haun2019why}. Again, this is wholly due to the internal connectivity of the complex (which has been brought about by adaptation to the environment), and not to some structure inherent in the stimulus.

In general, for perceptual richness to be high, several conditions must hold: there must be a complex capable of supporting a large repertoire of distinctions and relations (intrinsic meanings); the complex must be currently conscious, specifying a large \phistructure{}; it must be connected to the environment, that is, it must be ``awake,'' so that the stimulus can trigger an internal state through the ports-in; and its intrinsic connectivity must favor the percolation of the effects of the stimulus throughout the complex, so that many subsets contributing distinction \phifolds{} have high triggering coefficients.

Accordingly, regardless of how well a stimulus percolates through a system, nothing is experienced if its units are unable to support a \phistructure{} of high $\Phi$. This is clearly the case in deep slow wave sleep, when we lose consciousness due to neuronal bistability \autocite{tononi2008why}. Even though responses to sensory stimuli can be detected in many brain areas \autocite{yang2007effects}, we experience (next to) nothing. On the other hand, when dreaming, and under certain anesthetics, we can be highly conscious but severely disconnected \autocite{tononi2024consciousness}. This is due to neuromodulatory changes that prevent the exogenous triggering of states of the complex and instead promote its triggering by endogenous sources \autocite{aru2020apical}. The triggering coefficient allows one to discriminate between components of an experience/\phistructure{} that are triggered by stimuli (coefficient close to 1) and those that are imagined, hallucinated, or correspond to stimulus-independent thoughts or emotions (coefficient close to 0). Note also that triggering coefficients may be high for subsets of a system that are either outside the main complex or do not contribute distinctions and relations. For example, stimulus-triggered activity can be detected in brain regions that do not appear to contribute directly to experience \autocite{nakai2022representations}. Furthermore, subsets of the main complex whose cause-effects are not congruent with the cause-effect state specified by the complex as a whole would not contribute to experience \autocite{albantakis2023integratedb}. This may happen, for instance, during dreaming sleep, or under conditions of binocular rivalry, when non-dominant stimuli can trigger neuronal activity throughout the sensory hierarchy without contributing to experience \autocite{hesse2020newa}. It can also happen if a stimulus affects the complex at the wrong grain (finer or coarser; \cite{hoel2016can, marshall2018blackboxing}). Finally, the triggering coefficient can be used to determine how far a stimulus percolates through the complex and whether it is interpreted in a shallow or deep manner depending on the \phifolds{} it triggers.

\paragraph{Perception and representation.}~

It is often assumed that the brain forms “representations” of external entities and events by “processing information” conveyed by sensory stimuli.\footnote{Traditionally, representation can be defined as the formation of a mental image of the environment with input from memory and previous knowledge, preceded by sensation---the reception and initial encoding of stimuli---and perception---their categorization into meaningful concepts. In the present context, one could define sensation as the contribution to a \phistructure{}, weighted by the triggering coefficient, by the units of the complex that respond directly to features of the sensory interface (L1 ports-in, in the model); perception in a strict sense as the contribution by units responding to configurations of features (L2) and to categories of features  (L3); and representation (perception in a broad sense) as the portion of the \phistructure{} that refers to causal features in the environment (with the qualifications mentioned in the main text).} By IIT, the meaning of every content of experience is defined intrinsically by the corresponding \phifold{}. Therefore, not only is the meaning the same whether or not it is triggered by a stimulus, but it is also the same whether or not a percept “refers to” an actual entity or event in the environment. Moreover, even when a percept refers to some causal features of the environment, the mapping is typically not straightforward. This is because the environment is highly non-stationary, which may make it difficult to identify reliable referents, and because the meaning depends in a fundamental sense on a system’s intrinsic connectivity.

Consider again system B1 in \cref{matching:fig:representation}, which was designed to respond preferentially to stimuli typical of environment E1, which consists of one 3-segment generating process and one 2-segment generating process against a background of noise. This environment leads to stimuli that co-activate two or three contiguous units on the sensory interface, without activating other units in a detector's receptive field, more frequently than would be expected by chance. By design, these stimuli percolate all the way to L3 and trigger distinction \phifolds{} of high $\Phi{}$. As we have seen, from the intrinsic perspective of B1, the \phistructure{} triggered by such stimuli would be experienced as a “segment,” binding an invariant with a specific configuration and spatial location. In this case, the percept triggered in B1 might be considered as a “representation” that reflects causal features of E1—a “reference” to something “objective.” However, even in this simple case, the mapping between features of the environment and the structured meaning triggered by a stimulus may not be straightforward. Consider environment E1b, which contains two 2-segment generators and no 3-segment generator. When system B1 is exposed to environment E1b, the percept triggered by a stimulus containing 01110 (an apparent 3-segment) in system B1 will be that of a “segment,” even though in E1b only stimuli containing 01100 or 00110 (2-segments) occur frequently, while 01110 occurs at chance levels (conditional on the occurrence of 2-segments). Thus, there is nothing in the environment that the experience of a 3-segment can properly represent, beyond a coincidental occurrence of co-located 2-segments (or a 2-segment with a single unit activated by chance). Occasionally, even a stimulus from environment E3 may trigger a “segment” percept, even though E3 is a collection of independent noise sources. In fact, B1 will perceive any input from E3 as “spatially” extended, even though in E3 there is no “spatial” structure to be represented. On the other hand, in systems that have adapted to the same environment, one can expect that the perceptions triggered by typical stimuli are likely to represent many causal features of that environment that are relevant for survival, and that they will likely be similar across systems.

In short, a \phistructure{} triggered by a stimulus---a percept--- is always an interpretation based on its intrinsic meanings, but the interpretation does not always “represent” or “refer to” some causal feature of the environment, and if there is such a relation it is typically not straightforward.\footnote{This is partly in line with classic notions that perception is to some extent a construction (\eg \autocite{wertheimer1938laws, gregory1966eye}) or an inference (\eg \autocite{von1867handbuch, rock1983logic, lee2003hierarchical}).} Accordingly, by IIT the connection between an experience and the environment that triggers it is more akin to an \emph{adaptation}---an interpretation that works well enough---than to a representation in the conventional sense.

\paragraph{Perceptual differentiation and matching.}~

As we have seen, \emph{perceptual differentiation} is assessed by considering the set of intrinsic meanings triggered in a complex by a sequence of stimuli. For perceptual differentiation to be high, different stimuli in the sequence should not only trigger \phistructures{} with high perceptual richness, but also \phistructures{} that differ from each other, being composed of different distinctions and relations. The upper bound on perceptual differentiation is the overall \emph{intrinsic differentiation capacity} of the complex, corresponding to the diversity of the \phistructures{} that can be obtained by initializing the complex in all possible states.

\textit{Matching} was defined as the maximum expected difference between the perceptual differentiation triggered in a complex by stimulus sequences sampled from an environment, and that triggered by random stimulus sequences. The environment (or ``world'') is assumed to be much broader than the complex with respect to the number of units that constitute it and much deeper with respect to the length of the sequence of states it can visit. The sampling is mediated by the complex's sensory interface and is sensitive to its actions in the environment. As we have seen, taking the maximum difference ensures that matching captures the growth of perceptual differentiation with typical sequences, rather than trivially accruing with random ones.
Realistically, perceptual differentiation will be assessed with stimulus sequences much shorter than those yielding maximal values of matching. Therefore, it is essential to choose ``representative'' samples. These can be expected to yield optimal values of perceptual differentiation for any length of stimulus sequences.

In general, for matching to be high, the intrinsic meanings triggered in a complex by stimulus sequences sampled from its environment must ``resonate'' with features of the environment that preferentially generate those sequences, thereby differing from chance. A brain well adapted to its environment would be expected to allocate a large set of its intrinsic meanings to reflect causal features of the environment that are relevant for its survival. Typical stimuli from that environment should then trigger rich perceptual structures, and different stimuli should trigger different structures.

In the model, when system B1 is presented with sequences of stimuli from the `segment' environment E1, different stimuli activate different units not only at L1, but also different detector units at L2, each time triggering a \phistructure{} that differs markedly from the previous ones (\cref{matching:fig:matching}B, top). By contrast, random stimuli do not tend to trigger such high perceptual differentiation, resulting in a high matching value (\cref{matching:fig:matching}G). When B1 is presented with sequences from the `centered odd' environment E2, L2 units are activated less frequently, so the triggered \phistructures{} are not as rich and not as diverse (\cref{matching:fig:matching}E, top) and are similar to those triggered by noise, thus yielding a low matching value. Conversely, B2 shows higher perceptual differentiation when exposed to sequences from E2 in comparison to E1, and thus matches well to E2, and poorly to E1.

A model with additional high-level units responding to configurations and invariants such as dots and lines, if exposed to an environment with an abundance of appropriate stimuli, would show higher values of perceptual differentiation. The human brain, with its large repertoire of units selective for many different configurations and invariants, can be expected to show very high values of perceptual differentiation and matching in environments that reliably trigger those units.

Again, we can relate to this situation phenomenally. Watching a favorite movie triggers all kinds of experiential contents---different people and animals and plants, many different objects, countless different events---yielding high perceptual differentiation. The same movie temporally scrambled triggers fewer intrinsic meanings, as we fail to experience plot, suspense, and resolution. A random sequence of images from an unfamiliar domain, say, high-resolution scans of geological specimens, triggers even fewer meanings. Finally, “TV noise” triggers even less---a spatially extended flickering canvas and perhaps the concept of “noise”---despite being more differentiated, in an extrinsic, information-theoretic sense, than the frames of a film.

We saw above that, for a highly adapted brain, the percept triggered by stimulus sequences sampled from its environment should reflect relevant features of that environment. To the extent that is true, the percept triggered by a single stimulus might be considered as a representation. The same logic applies to perceptual differentiation and matching: the different \phistructures{} triggered by representative sequences of stimuli will reflect many different causal features of the environment that are likely relevant for survival. The reason is that, over the time scales of evolution, development, and learning, the intrinsic connectivity of the brain will have been molded by plastic interactions with that environment. We can thus consider perceptual differentiation triggered by stimulus sequences sampled from its environment not only as a measure of the expected “meaningfulness” of that environment to that brain, but also as an estimate of the “matching” between intrinsic meanings and extrinsic causal processes typical of that environment \autocite{tononi1996complexity, tononi2012integrated}. However, just as the notion of representation must be qualified, so must that of matching. Intrinsic meanings triggered by sequences of stimuli do not always “represent” or “refer” to causal features of its environment, and if they do, it is typically not in a straightforward manner. Even so, one can investigate which features of the environment contribute most to matching by measuring differentiation using different sequences of stimuli. For example, scrambling the sequence of stimuli highlights how differentiation depends on causal processes in the environment that are responsible for the temporal structure of stimuli. In fact, measuring perceptual differentiation after scrambling stimulus sequences in various ways can be employed to systematically investigate various aspects of meaningfulness. For instance, with respect to text, one could scramble the order of paragraphs, that of sentences, that of words, and that of letters, and obtain estimates of how intrinsic meaningfulness is affected at each step.

On this basis, one can expect that brains should match especially well, and with greater inter-subjective agreement, certain causal features of a shared environment. These are features that are either ubiquitous, such as the smoothness of the external world in “space” and “time,” or highly clustered because they originate in external entities that “hang together well” at a scale relevant for adaptive interactions, such as animals, plants, and so on. Again, however, perceptual differentiation will be non-zero for features that are highly idiosyncratic and whose external referent may be highly disjunctive, such as “beauty.”

\paragraph{Interpretive and generative power.}~

A brain with high values of perceptual differentiation can provide a rich interpretation of an environment in response to many different stimuli. Under adaptive constraints, those interpretations will typically refer to various causal properties that are true of its environment---they will reflect the matching between intrinsic meanings and extrinsic causal processes. High perceptual differentiation then implies high \emph{interpretive power}. As we have seen, perceptual differentiation can only be high if intrinsic differentiation capacity is high to begin with, which requires that many different \phistructures{} have high values of structure integrated information ($\Phi{}$). High $\Phi{}$, in turn, requires a complex that is highly integrated: every part of it must be able to interact bidirectionally with the rest of the complex \autocite{albantakis2023integratedb}. In a sensory hierarchy, this implies that the intrinsic connectivity (whether considered bottom-up, top-down, or lateral) must ensure that activating or deactivating subsets of units anywhere in the complex should make a difference elsewhere.

A consequence of this requirement is that a complex with high matching to a given environment does not just possess high interpretive power, but likely also high \emph{generative power}. This can be understood as the ability of the complex to produce, through its intrinsic connectivity, sequences of intrinsic states that are similar to those observed when it is connected to the environment. For example, consider the organization of posterior-central cortex as a hierarchical lattice poised to be triggered by stimuli via bottom-up input. Being maximally integrated, the same lattice can also enable, under the right neuromodulatory conditions, the endogenous triggering of activity patterns independent of the state of primary cortical areas. This is what seems to be happening during dreaming, imagination, or mind-wandering, when the corticothalamic system generates endogenously activity patterns similar to those triggered by environmental stimuli \autocite{tononi2024consciousness}. Regardless of whether these activity patterns are triggered exogenously and bottom-up (from the particular to the general) or endogenously and possibly top-down (from the general to the particular), the underlying intrinsic connectivity is the same. Therefore the associated \phistructure{}---which is to say, the corresponding intrinsic feeling/meaning---will be just as similar.

Dreams provide the most obvious demonstration that the substrate of consciousness has both interpretive and generative power. They reflect a large amount of internalized constraints about causal processes in our environment: we dream of humans, animals, plants, and objects, faces and places, colors and sounds---not of things that we are unequipped to experience, in principle, during wakefulness. In short, dreams reflect the interpretive power of the substrate of consciousness. But dreams also demonstrate the substrate's generative side, because we can evidently bring into existence world-like experiences endogenously, without the need to trigger them exogenously \autocite{nir2010dreaming,pearson2019human}. Imagination during wakefulness also testifies to our ability to trigger different experiences endogenously, although the vividness and detail of imagination vary greatly among individuals \autocite{tononi2024consciousness}. This generative power thus frees the organism from the ``tyranny of the here and now'' and allows it to plan ahead and try out imaginary scenarios. It is also essential for allowing endogenous reactivation during sleep to promote matching, as will be discussed below.

\paragraph{The role of wakefulness and sleep in increasing and preserving matching.}~

A good matching between intrinsic meaning and causal features of the environment has obvious adaptive significance. But achieving and preserving high matching is not trivial, not only because of the richness, diversity, and non-stationarity of the environment, but also because the brain is mostly connected to itself, and only a comparatively small subset of its neurons function as ports-in (and ports-out) in its interactions with the world.

As argued elsewhere \autocite{tononi2012integrated, hashmi2013sleepdependent, tononi2014sleep}, a critical role in promoting matching is likely played by the repeated alternation between periods of synaptic up-selection during waking, when the brain is connected to the environment, and synaptic down-selection during sleep, when it is disconnected but spontaneously active.

As argued above, at the time scale of experience (tens to hundreds of milliseconds), stimuli trigger activity patterns in the brain that are endorsed by the intrinsic connectivity through fast, dynamic changes in the efficacy of interactions. However, the same activity patterns can also initiate long-lasting changes. During wakefulness, these changes are primarily manifested as a net potentiation of synaptic strength \autocite{tononi2020sleep}.\footnote{Again, this is because, owing to energy constraints, strong activation should be reserved for the selective signaling of highly meaningful stimuli that need to percolate deeply within the system \autocite{balduzzi2013what}.} Bottom-up, top-down, and lateral connections that close cause-effect loops within the brain may be especially likely to be potentiated \autocite{tononi2014sleep, hashmi2013sleepdependent}. For example, if an activity pattern triggered by a stimulus through bottom-up connections is congruent with the effects of top-down connections, then all those connections are potentiated, up-selecting the distinctions and relations involved. During development and learning this process can be extremely complex, relying on interleaved timelines of synaptic growth and pruning, spontaneous activity, regularities in low-level input patterns, the active sampling of the environment by the organism, and neuromodulatory systems that gate plasticity based on salience and rewards.\footnote{As illustrated in \autocite{nere2013sleepdependent, hashmi2013sleepdependent}, this can be done by strengthening clusters of feedforward connections carrying suspicious coincidences of firing (primarily through AMPA receptors) preferentially when “endorsed” by the firing of feedback connections targeting the same dendritic domain (primarily through NMDA receptors). The endorsement through feedback indicates to a neuron that through its effects it “made a difference” downstream through loops involving some brain circuit and potentially the environment (perception-action loops, \textcite{hashmi2013sleepdependent}). Also, because the same neuron can participate in different subsets—each specialized for different suspicious coincidences—this process can yield a large repertoire of high-order mechanisms, each specifying a different \phifold{}. It also ensures that the brain learns to generate intrinsic meanings that resemble those triggered by environmental stimuli, giving it the ability to predict, imagine, and plan.}

Initially, synaptic up-selection will increase matching by increasing the selectivity and percolation of responses to typical stimuli from the environment. Over time, however, synaptic potentiation can lead to catastrophic consequences \autocite{tononi2014sleep}. First, the brain will inevitably up-select spurious coincidences of firing that do not reflect genuine causal processes in the environment. Moreover, many coincidences of firing will be triggered by intrinsic activity, rather than by extrinsic inputs, and would soon dominate plasticity. This would progressively reduce the degree to which intrinsic meanings refer to causal processes in the environment. An unchecked increase in synaptic strength would also decrease the selectivity of neuronal responses, saturate neurons’ ability to learn new coincidences, and impose a major burden on cellular homeostasis \autocite{tononi2014sleep}.

For these reasons, as argued by the synaptic homeostasis hypothesis (SHY) of sleep function \autocite{tononi2020sleep}, neurons need periods in which they are disconnected from the environment (off-line) and can undergo synaptic down-selection while sampling comprehensively activity patterns generated intrinsically. In this process, synapses that were strengthened owing to spurious coincidences in the environment or to internal “fantasies,” can be weakened systematically, night after night. Instead, coincidences reflecting causal processes in the environment can be strengthened systematically, day after day. In this way, intrinsic meanings that refer to environmental regularities are enhanced or preserved, whereas statistical noise and fantasies are averaged away. The repeated cycling between synaptic up-selection during wakefulness, guided by environmental regularities, and comprehensive synaptic down-selection during sleep, can promote perceptual differentiation while maintaining flexibility in the face of environmental changes \autocite{hashmi2013sleepdependent,nere2013sleepdependent}. Moreover, response selectivity and specificity are restored, learning ability is de-saturated, and cellular stress is reduced \autocite{tononi2014sleep}.

\paragraph{Assessing differentiation and matching with neurophysiological data.}~

Unfolding \phistructures{} systematically to evaluate perceptual differentiation is only feasible for extremely simple substrates such as the model employed here. However, it is possible to obtain practical approximations using \emph{neurophysiological differentiation} as a proxy \autocite{mensen2018differentiation,boly2015stimulus,mensen2017eeg,mayner2022measuring,gandhi2023survey}. The reason is that activity patterns over a complex fully determine the associated \phistructures{}, as long as we can make reasonable assumptions about the border and grain of the substrate of consciousness, the mechanisms of the units constituting it, and its relative stability. Thus, even a coarse estimate of neurophysiological differentiation may be adequate to rank relative levels of perceptual differentiation for brain regions with the right connectivity for supporting consciousness.

For example, we previously showed using fMRI and estimates of Lempel-Ziv complexity that neurophysiological differentiation within cortex (especially its posterior-central regions) was higher for a movie than for an equivalent sequence of TV noise frames, and that it was higher if the movie was in the proper sequence rather than scrambled \autocite{boly2015stimulus}. Furthermore, we showed using high-density EEG that an estimate of neurophysiological differentiation was higher for sets of stimuli or movie clips that the subjects found more meaningful \autocite{mensen2018differentiation}. Similar results were obtained at cellular resolution with calcium imaging \autocite{mayner2022measuring} and Neuropixels recordings \autocite{gandhi2023survey} in mice. On the other hand, measuring neurophysiological differentiation over the retina (or the cerebellum) would say nothing about perceptual differentiation, because these systems cannot support consciousness due to their internal organization \autocite{albantakis2023integratedb}. Similarly, measuring differentiation at a spatial or temporal grain that is either too fine or too coarse could produce misleading results \autocite{hoel2016can,marshall2018blackboxing}.

In future work, measures of neurophysiological differentiation could be employed to monitor brain development and the effectiveness of learning in neurotypical individuals. Or they could be used to infer whether individuals that have lost the ability to communicate may nonetheless retain the capacity to meaningfully interpret sensory stimuli. They could also help in assessing which stimulus sequences are most meaningful for subjects who are not neurotypical, such as individuals with autism or schizophrenia. In all cases, measures of differentiation should be obtained with careful consideration of the likely location of the substrate of consciousness, while ensuring that subjects remain behaviorally engaged and therefore connected. Stimulus sequences should be chosen such that they are as representative as possible. Stimuli should also be highly relevant for the subjects of study. Estimates of neurophysiological differentiation could then be employed to compare the meaningfulness of different sequences to the same individual, say, by progressively adding or removing contents, statistical regularities, and scrambling stimuli. In principle, they could also be employed to evaluate the relative meaningfulness of different stimulus sequences for species whose ecological habitat and brain organization is not well known. In such cases, a search strategy aimed at maximizing differentiation might be used to infer what might be meaningful to that species. This point is important conceptually because it illustrates how optimizing differentiation can establish what aspects of an environment are meaningful to an organism without presupposing or predefining what those aspects might be.

Maximizing neurophysiological differentiation using a large set of diverse stimuli likely to be meaningful to a subject would also help in estimating a complex's \emph{intrinsic differentiation capacity}. Contingent on various assumptions about the border and grain of a complex and the mechanisms of its constituent units, its intrinsic differentiation capacity should be proportional to the $\Phi$ value of the \phistructure{} it specifies when in a typical state (see \textcite{marshall2016integrated}). In other words, maximal neurophysiological differentiation could also serve as an estimate for the quantity of consciousness ($\Phi$).

\paragraph{Intrinsic and extrinsic approaches to meaning and information.}~

According to IIT, as we have seen, an experience is a structure of maximally irreducible cause-effect power---a \phistructure{} composed of integrated information---rather than a code, a computation, a function, or a process \autocite{tononi2014sleep,zaeemzadeh2024shannon}. An experience exists “for itself,” rather than with reference to something else, and it exists “here and now,” rather than by reference to some past or future event. Its intrinsic meaning, or “information content,” is the same as its feeling: it is identical to a \phistructure{}. In this sense, meaningful information is conscious information, and communicating meaningful information ultimately requires the triggering of similar \phistructures{} among conscious subjects \autocite{zaeemzadeh2024shannon}.

IIT’s approach contrasts with common views that take an extrinsic perspective on information and meaning. For example, the brain is often portrayed as an ``information processing'' device, where information is conceptualized along the lines of Shannon's information theory \autocite{zaeemzadeh2024shannon}. Accordingly, the brain would decode information provided by a stimulus, understood as a message or symbol, and compute appropriate outputs based on that information and on additional information provided by memories and goals. Top-down ``priors'' are thought to act as error correction codes, carrying ``contextual'' information to properly decode or fill in noisy or incomplete stimuli, eventually yielding an activity pattern that is a better ``code'' for the meaning of the input. It has also been suggested that codes or messages that are broadcast globally through long-range connections correspond to ``conscious processing'' \autocite{dehaene2011global}. Alternatively, the brain has been portrayed as a ``predictive processing'' device that implements a reverse message-passing scheme, where top-down inferences are updated based on error signals provided by stimulus information \autocite{cao2020new}.

However, as explicitly recognized by Shannon, information theory is concerned with the optimal encoding, transmission, decoding, and storage of symbols, and has nothing to say about the symbols' meaning \autocite{shannon1948mathematical}. Therefore, it is not clear what would determine the meaning of the activity patterns or codes resulting from such processing \autocite{brette2019coding, buzsaki2019brain}. It is tempting to refer such meanings backwards, to sources in the environment, or forward, onto actions or at least to intermediary computations within the brain. More generally, computational/functionalist approaches equate meaning to a computation or function being performed with respect to inputs, outputs, and internal states, such as memories and goals \autocite{putnam1973meaning}. But while these approaches can account for \emph{what a system does}, from the extrinsic perspective of a conscious observer, they do not account for what the system \emph{is like} from its own intrinsic perspective.\footnote{IIT’s intrinsic view of the meaning of an experience as identical to a \phistructure{} also differs from extrinsic views that assign meaning based on the similarities among activity patterns \autocite{kriegeskorte2008representational} or their location within ``conceptual spaces'' \autocite{gardenfors2000conceptual}.}

\paragraph{Intrinsic and extrinsic matching.}~

Finally, a key implication of IIT concerns computers implementing artificial intelligence (as well as computers potentially simulating our brains). According to IIT, computers are not well-suited to support \phistructures{} of high $\Phi$ because of their internal organization, which is highly modular and typically feed-forward \autocite{findlay2024dissociating}. Yet such computers may soon become functionally equivalent to us---navigating the environment, answering questions, and pursuing seemingly goal-directed behaviors. From a human vantage point, they would behave \emph{as if} they were perceiving and understanding the world just as we do. However, they would not experience what we experience---nothing we feel would feel like anything or mean anything to them---and they would have neither intrinsic goals nor free will \autocite{tononi2022only}. Nevertheless, extrinsic measures of differentiation/matching could be defined by analogy with intrinsic measures. For example, one could assess the extrinsic meaningfulness of different environments or stimulus sequences by measuring the mutual information between sensory interfaces and different subsets of specialized ``hidden'' units (see also \textcite{tononi1996complexity}).

\paragraph{Future extensions.}~

In future work, IIT's approach to intrinsic meaning could be extended in several directions, some of which will be mentioned briefly. A natural extension is to evaluate connectedness and triggering coefficients on the output side---the motor interface. Just as an input over a sensory interface is connected to the main complex if it is has an actual effect on the state of the complex's subsets, an output is connected to the main complex if the latter has actual effects on a motor interface. Accordingly, just as an input leads to a percept to the extent that it triggers distinction \phifolds{} within the \phistructure{} of the main complex, an action is intended to the extent that a \phifold{} specifies effects that trigger that output. And just as a distinction \phifold{} can include high-level concepts that represent objects and events, an intention \phifold{} may include high-level concepts that establish goals and action plans. Because of its ability to control its outputs (say, eye or hand movements), a complex can also change the way it samples its inputs and thus influence its own responses to the environment. Active exploration of the environment is important not only for refining representations here and now, but also during learning, when it can greatly enhance our ability to sample causal features of the environment (\eg, while a passive observer could not distinguish between the presence of two partially overlapping 2-segments and a real 3-segment in \cref{matching:fig:representation}, an actor could, thus allowing it to ``prune'' its superfluous ``concept cell'' sensitive to 3-segments). Furthermore, actions can modify the environment and create new things or processes, some of which may match “invented” intrinsic meanings. Said otherwise, systems adapt not only by matching the environment on the input side, but also by \emph{shaping} it on the output side.

A further extension will be to apply IIT's causal powers analysis to characterize causal entities (metastable clusters of causal features) and processes (successions of overlapping causes and effects) as sources of regularities that organisms sample from the environment \autocite{tononi2022only}. This will help to ground the notion of matching to regularities in the environment, and to relate subjective representations to objective (or at least inter-subjective) \textit{knowledge} \autocite{tononiforthcomingbeing}. In practice, complexes with an intrinsic connectivity capable of specifying a large repertoire of distinctions and relations come about because of adaptive interactions with a rich environment, as suggested by computer simulations using genetic algorithms \autocite{albantakis2014evolution}. As shown there, a complex of high $\Phi$ can have an adaptive advantage over less integrated systems (say, feedforward only or modular) because it can “pack” a larger number of functions over fewer elements, which is critical under biological constraints on unit numbers and energy supplies. As shown here, given a complex of high $\Phi$, a simple, unstructured stimulus can trigger a cause-effect structure that synthesizes, “here and now,” a long history of adaptations to countless causal features of the environment.\footnote{ As pointed out in \textcite{tononi2012integrated}, a complex of high $\Phi$ has additional advantages with respect to learning, internal routing, access/control, and understanding, thanks to its compositional and relational architecture.} In this sense, perception truly goes “beyond the information given” \autocite{bruner1973information}.\footnote{The triggering of a large cause-effect structure should not be thought of as a “memory” in the sense of a recollection of a past event, but as interpretation in the light of past adaptations.}

Finally, another extension of the present framework will be the exploration of the bounds of intrinsic meaning. In principle, one could attempt to build systems having maximal intrinsic differentiation capacity \autocite{zaeemzadeh2023upper} with no regard for adaptive constraints but optimally suited to interpreting their own states. However, even a system optimized in this way would not be able to specify intrinsically the immense number of distinct meanings that could be specified over any of its states (not even remotely; see \textcite{feldman2003catalog}). Therefore, in one relevant sense, it would not be able to fully understand itself.\footnote{The incompleteness of the intrinsic meanings that can be specified by a system over itself through its causal mechanisms bears a resemblance to the formal incompleteness of logical systems demonstrated by G\"odel \autocite{goedel1931ueber,raatikainen2022godels}.}

\section*{Acknowledgments}
\label{matching:sec:acknowledgements}
We thank Larissa Albantakis and William Marshall for their contributions during the early stages of this project; Leonardo S.\ Barbosa, Melanie Boly, Tom Bugnon, Keiko Fujii, Andrew Haun, Armand Mensen, and Shuntaro Sasai for helpful discussions; and Larissa Albantakis, Chiara Cirelli, Francesco Ellia, Graham Findlay, Matteo Grasso, and Alireza Zaeemzadeh for valuable comments on the manuscript. B.E.J. was supported in part by the Norwegian Research Council (FRIPRO grant no. 335828). G.T. acknowledges support from Templeton World Charity Foundation (nos. TWCF0216 and TWCF0526). The opinions expressed in this publication are those of the authors and do not necessarily reflect the views of Templeton World Charity Foundation.

%
%
\bibliography{references.bib}

\section*{S1 Text}
\label{matching:sec:supporting-information}
\label{matching:appendix}

\subsection*{The substrate model and its units}
\label{matching:appendix:model}

The two substrate models used in this work each comprise 21 units, organized in
4 hierarchical levels: a sensory interface of 8 units, a lattice level of 8
units, a configuration detector level of 4 units, and an invariant
level of 1 unit. Each level has an associated activation function, shared by each unit in that level, so that within a level the units are mechanistically identical and differ only in the identity of their inputs and outputs (\cref{matching:fig:model}A). The substrates were built using the \texttt{substrate\_modeler} package
available on GitHub at \url{https://github.com/bjorneju/substrate\_modeler.git}. Here, we give a
narrative description of their functionality. 

Although the units are inspired by neural mechanisms, they are not models of individual neurons. Rather, they are more analogous to small neural circuits consisting of both inhibitory and excitatory neurons, \ie they are `macro' units \autocite{hoel2016can,marshall2018blackboxing}.
Furthermore, the time scale of a single dynamical update in the model is a `macro' time scale long enough to allow the constituent parts of each unit to adapt to the input before the resulting state is produced and transmitted to the unit's outputs.
Thus, the activation functions are state-dependent, both with respect to the state of the unit itself and the state of its inputs.
Finally, the final transmitted state of each unit is binary: it is `ON' if the `micro' unit corresponding to its output reaches a certain activity threshold (analogous to \eg bursting of primary cells) and `OFF' otherwise.

\subsubsection*{$\mathbf{\sensoryinterface}$: Sensory interface units}
\label{matching:sec:sensory-surface-units}

Each unit in the sensory interface receives a single input from itself, and its state is determined by a
sigmoidal activation function. The details of the
activation function are irrelevant as these units are always clamped to a particular stimulus in our analyses; it is defined explicitly only to include the sensory interface units in the TPM of the substrate.

\subsubsection*{L1: Lattice units}
\label{matching:sec:lattice-units}

Each unit $k$ in the lattice level receives a single bottom-up input $x_k$ from the sensory interface. In addition, each lattice unit receives inputs $I$ from within the system: one from itself and two from its nearest neighbors within the lattice level (or only one if $k$ is on the boundary, \ie, units $A$ and $H$), and a single top-down input from the top-level unit $M$. The activation function is a combination of two sub-functions.

The first function $f_1(x_k, s_k)$ compares the unit's current state $s_k$ with the bottom-up input $x_k$. If the input state differs from the unit's current state, the unit is driven to flip its state.
This means that a lattice unit in its default state (`OFF') will be reliably activated if the bottom-up input is `ON', and will reliably turn `OFF' again if the bottom-up input ceases.

The second function determines how the state of the lattice unit is influenced by lateral, self, and top-down input from the top-level unit. It is implemented as a sigmoid function of an input state $\mathbf{I^*}$, the shape of which is parameterized by the connection weight $w_k$, the current state of the unit $s_k$, and current state of its inputs $\mathbf{I}$:
\begin{equation}
    \sigma(\mathbf{I^*}; s_k, \mathbf{w_k}, \mathbf{I}) \;=\;
    \frac{
        1
    }{
        (1 + \exp\left[
            -s_k \sum_{j=1}^{\card{\mathbf{I}}} I_j\,g(I_j,s_k)\,w_{k,j} I_j^*
        \right])
    }.
\end{equation}
Here an `OFF' state is counted as $-1$ and an `ON' state is counted as $1$, similar to the Ising model.

The factor $g(I_j,s_k)$ modifies the coupling strength of the input from unit $j$ to unit $k$ and is defined as follows:
\begin{align}
g(I_j, s_k) \;=\;
\left\{
    \begin{aligned}
         &1.5  &&\text{if}\ I_j = \text{ON,}  &&s_k = \text{ON} \\
         &1    &&\text{if}\ I_j = \text{OFF,} &&s_k = \text{OFF} \\
         &0.75 &&\text{if}\ I_j = \text{ON,}  &&s_k = \text{OFF} \\
         &0.5  &&\text{if}\ I_j = \text{OFF,} &&s_k = \text{ON.}  \\
    \end{aligned}
\right.
\end{align}

The state-dependent sigmoid function $\sigma(\mathbf{I^*}; s_k, \mathbf{w_k}, \mathbf{I})$ is intended to implement some of the complexities of neural microcircuits as a single formula. It accounts for the adaptive and state-dependent nature of both coupling strengths and neural mechanisms by implementing a form of short-term plasticity via the $g$ factor. The multiplication by $s_k$ in the formula ensures that the connections to a unit that is `ON' ($+1$) are effectively excitatory, while connections to a unit that is `OFF' ($-1$) are effectively inhibitory. Thus, the function explicitly incorporates the reinforcement of the unit's own state by adjusting the net strength and effective sign of the inputs to the unit.

The functions $f_1(x_k, s_k)$ and $\sigma(\mathbf{I^*}; s_k, \mathbf{w_k}, \mathbf{I})$ are combined to obtain the probability of activation by taking the `maximally selective' one, \ie the one that deviates maximally from chance:
\begin{equation}
    \Pr(k = \text{ON}) \;=\; \argmax_{p \,\in\, \left\{f_1(x_k, s_k),\; \sigma(\mathbf{I^*}; s_k, \mathbf{w_k}, \mathbf{I})\right\}} \card{p - 0.5}.
\end{equation}
This ensures that the states of the lattice units are reliably driven by the bottom-up input from the sensory surface, while also allowing the units to ``endorse'' their current state whenever they are correctly set by the input.

\subsubsection*{L2: Configuration detectors}
\label{matching:sec:configuration-detectors}

Each unit $d$ in the configuration detector level receives bottom-up input from a unique tuple $X$ of five lattice units (\eg, unit $I$ receives inputs from $(A,B,C,D,E)$).
Like the lattice units, each configuration detector also has a self-connection and receives a top-down input from the top level unit $M$. The activation function is also a combination of two sub-functions.

The first function implements a high probability of activation in response to a particular set of input configurations. The two substrates used in our analyses (B1, the `segment' detector, and B2, the `centered odd' detector) differ only in the set of configurations that this function is selective for:
\begin{equation}
f_2^{B1}(X) \;=\;
    \begin{cases}
        0.99 &\text{if}\ X = \{(0,1,1,1,0), (0,1,1,0,0), (0,0,1,1,0)\} \\
        0.01 &\text{otherwise}.
    \end{cases}
\end{equation}
\begin{equation}
f_2^{B2}(X) \;=\;
    \begin{cases}
        0.99 &\text{if}\ X = \{(0,1,1,1,0), (0,0,1,0,0)\}  \\
        0.01 &\text{otherwise}.
    \end{cases}
\end{equation}

Similar to the lattice units, the second function is a state-dependent sigmoid function of the self-connection and top-down input, $\sigma(\mathbf{I^*}; s_k, \mathbf{w_k}, \mathbf{I})$.

The probability of activation is then obtained by combining the functions `in series':
\begin{equation}
\label{matching:eq:configuration-detector}
    \Pr(d = \text{ON}) \;=\; f_2(X) + ( 1 - f_2(X) )\, \sigma(\mathbf{I^*}; s_k, \mathbf{w_k}, \mathbf{I}).
\end{equation}
Thus, these units are always counterfactually dependent on the state of lattice units, while also allowing the units to endorse their own state.

\subsubsection*{L3: Invariant unit}
\label{matching:sec:segment-detector-unit}

The invariant unit at the top of the hierarchy works similarly to the configuration detectors, except that it receives bottom-up input $X$ from the four configuration detectors and intra-level input $I$ from itself. Its bottom-up function $f_3$ activates strongly when at least one of the detectors is `ON':
\begin{equation}
    f_3(X) \;=\;
    \begin{cases}
        0.99 & \text{if~}\ {\displaystyle \sum_{x\in X} x \geq 1} \\
        0.01 & \text{otherwise}.
    \end{cases}
\end{equation}
The activation function is then defined as in \eqref{matching:eq:configuration-detector}, substituting $f_3$ in place of $f_2$.

\subsection*{Distinctions and relations}
\label{matching:appendix:distinctions-and-relations}

Our chief concern in designing the model system was to ensure it had enough recognizable functionality to allow a useful illustration of the formalism, which resulted in a system of 13 units (the 8 sensory interface units are not part of the system). The exponential time complexity of the IIT analysis makes exhaustive unfolding of the entire \phistructure{} impractical in a system of this size.
Consequently, we chose a representative sample of distinctions and relations to compute.
We chose mechanisms that feature connectivity motifs of interest based on prior work on how IIT may account for the experience of visual space \autocite{haun2019why} and ongoing work on how IIT may account for the experience of objects \autocite{grassohow}. We further restricted the mechanisms to those whose units are directly connected to each other.

Specifically, we chose to compute the distinctions specified by the following subsets of units within the system:
\begin{enumerate}[labelindent=\parindent, leftmargin=*, itemsep=-3pt, label=(\arabic*)]
    \item all subsets of size 1 (first-order mechanisms);
    \item all contiguous subsets of lattice units in level 1;
    \item all subsets of the level 2 configuration detectors; and
    \item subsets up to and including size 3 that span one or more levels.
\end{enumerate}

Furthermore, for each of the mechanisms, we restricted the set of possible purviews considered in the distinction calculation. We excluded any units not directly connected to the mechanism. We further restricted purviews to those that were likely to yield maximal $\phi_d$ for the mechanism, which was determined by consideration of the activation functions involved and a preliminary exhaustive search over all possible purviews for certain example mechanisms from each class in a representative selection of states.

Since the number of relations grows much faster than the number of distinctions, we computed only relations up to and including degree 3 (\ie, relations among up to three distinctions).

\end{document}